\documentclass[pre,twocolumn,nobibnotes,a4paper,secnumarabic,nofootinbib,floatfix,]{revtex4-1}
\usepackage{textcomp}
\usepackage{amssymb,amsmath,MnSymbol,ifsym,latexsym,bm}
\usepackage{wrapfig,subfigure}
\usepackage[Euler]{upgreek}
\usepackage{longtable, supertabular}
\usepackage{multirow}
\usepackage{float}
\usepackage{rotating}

\usepackage{setspace}
\usepackage{color}

\bibpunct[, ]{[}{]}{;}{n}{,}{,}

\newcommand{\Pe}{\mathrm{Pe}}
\newcommand{\kBT}{k_\mathrm{B}\,T}

\begin{document}
\title{Significance of thermal fluctuations and hydrodynamic interactions in receptor-ligand mediated adhesive dynamics of a spherical particle in wall bound shear flow}
\author{K. V. Ramesh}
\affiliation{IITB-Monash Research Academy, Indian Institute of Technology, Mumbai, India}
\altaffiliation{Department of Chemical Engineering,  Indian Institute of Technology, Mumbai, India}
\altaffiliation{Department of Mechanical \& Aerospace Engineering, Monash University, Clayton, Australia}
\author{R. Thaokar}
\affiliation{Department of Chemical Engineering,  Indian Institute of Technology, Mumbai, India}
\author{J. Ravi Prakash}
\affiliation{Department of Chemical Engineering, Monash University, Clayton, Australia}
\author{R. Prabhakar}
\email{prabhakar.ranganathan@monash.edu}
\affiliation{Department of Mechanical \& Aerospace Engineering, Monash University, Clayton, Australia}
\date{\today}

\begin{abstract}

The dynamics of adhesion of a spherical micro-particle to a ligand-coated wall, in shear flow, is studied using a Langevin equation that accounts for thermal fluctuations, hydrodynamic interactions and  adhesive interactions. Contrary to the conventional assumption that thermal fluctuations play a negligible role at high P\'{e}clet numbers, we find that for particles with low surface densities of receptors, rotational diffusion caused by fluctuations about the flow and gradient directions aids in bond formation, leading to significantly greater adhesion on average, compared to simulations where thermal fluctuations are completely ignored. The role of wall hydrodynamic interactions on the steady state motion of a particle, when the particle is close to the wall, has also been explored. At high P\'{e}clet numbers, the shear induced force that arises due to the stresslet part of the Stokes dipole, plays a dominant role, reducing the particle velocity significantly, and affecting the states of motion of the 
particle. The coupling between the translational and rotational degrees of freedom of the particle, brought about by the presence of hydrodynamic interactions, is found to have no influence on the binding dynamics. On the other hand, the drag coefficient,  which depends on the distance of the particle from the wall, plays a crucial role at low rates of bond formation. A significant difference in the effect of both the shear force and the position dependent drag force, on the states of motion of the particle, is observed when the P\'{e}clet number is small.  
\end{abstract}

\maketitle

\section{\label{s:intro} Introduction}

Particle-substrate adhesion is ubiquitous in biological systems. In the case of cell-surface adhesion, bonding between surface receptors expressed on the cell and complementary ligands expressed on any membrane or cell surface, allow the cell and the surface to specifically adhere to each other like a lock and a key, causing cell adhesion to be highly specific. Some examples are: binding of white  blood cells (leukocytes) to particular tissues by specific receptor interactions with surface ligands \citep{Springer1994301}; enhanced adhesion of infected red blood cells to artery and capillary walls due to surface receptors expressed by the malarial parasite, \textit{Plasmodium falciparum}~\citep{Cooke1994,Roy2005,Antia2007}; modulation of adhesive properties of cancer cells leading to  metastases from the primary tumor through the circulatory system~\citep{Cristofanilli2004,Bono2008}; viral docking to cell surface receptors~\citep{Alberts2002} \textit{etc.} \textit{In vitro} experimental techniques such as 
surface force apparatus, dynamic force spectroscopy, flow chamber experiments and optical trap force spectroscopy have been used extensively to probe various steps in microparticle adhesion.

Modeling and computer simulations further allow one to explore adhesion of particles coated with receptors moving near a ligand-coated substrate under the action of external forces~\citep{Pawar2008,Jadhav2007,Korn2007,Korn2008,English20043359, Hammer1992}. Such simulations are now being increasingly used to extract parameters such as rate constants in binding kinetics, binding energies and force constants \textit{etc.} by fitting model predictions through experimental data~\citep{Cristofanilli2004}. Currently, such studies use the extracted parameters for qualitatively comparing the behaviour of particles with different binding properties \textit{e.g.} healthy and malaria-infected red-blood cells have different adhesins expressed on their surface. The wide interest in drug delivery however suggests that quantitative accuracy may also be desirable for designing particles with specific adhesion targets. There are significant differences however between simulation approaches used in studies so far, 
particularly in relation to the treatment of thermal fluctuations and hydrodynamic interactions (HI) with the rigid ligand-coated substrate (the ``wall"). Our aim is to better understand the role played by these phenomena in determining the states of motion and the dynamics of an adhesive microparticle in a shear flow cell, so that a more judicious choice of simplifying assumptions and simulation approaches can be made depending on one's end goals.

The P\'{e}clet number $\Pe$ estimates the relative importance of the kinematics enforced by the imposed flow over the Brownian motion caused by thermal fluctuations. Typical shear rates at which shear-cell studies are carried out are around 1-100 s$^{-1}$. Such shear gradients correspond to very large $\Pe$ values for microparticles such as cells of size 1--100 $\mu$m in an aqueous medium.  Typically a high value of $\Pe \gg 1$ is assumed to indicate that thermal fluctuations are unimportant, and early simulations of particle adhesion~\citep{Hammer1992} ignored thermal fluctuations completely for this reason.  They demonstrated that the coupling of the shear flow with the on- and off-kinetics of the receptor-ligand interactions led to various states of motion of an adhesive particle at a wall, such as rolling, firm-adhesion, free diffusion, \textit{etc}. \cite{Korn2007,Korn2008} studied these states of motion at high $\Pe$ values by including Brownian fluctuations. They argued that although thermal 
fluctuations under such conditions are unimportant for the motion of the particle in the plane of the shear flow, they affect the orientational diffusion of the particle about the gradient direction, bringing receptors into proximity with ligands. This raises two questions: firstly, what if any are the changes to the states of motion  when thermal fluctuations are included in high-$\Pe$ simulations, and secondly, under what conditions are fluctuations unimportant and hence may be safely neglected in order to design less computationally-intensive simulations (i.e., relatively fast deterministic algorithms).

The adhesion models in the studies cited above incorporate hydrodynamic interactions of the spherical particle with  the wall~\citep{Hammer1992,Korn2007,Korn2008,Cichocki1998273}.  On the other hand, the Brownian Adhesive Dynamics (BRAD) algorithm of~\citet{English20043359} ignores HI completely to achieve a simpler set of Langevin equations. Wall HI has three principal effects. It causes the friction coefficient to diverge strongly as the wall is approached. Secondly, it leads to significant coupling between the translational and rotational degrees of freedom. These two effects give rise to a position dependent mobility matrix for the particle motion. In addition, the stress distribution on the particle surface induces a net hydrodynamic dipole which causes a flow-induced shear-rate dependent force.   The hydrodynamic simplicity of the BRAD algorithm allows for a greater focus on modeling of receptor-ligand interactions, but it is not clear if the net result of HI on dynamical states in a shear flow can be 
modeled by a simple renormalization of the friction coefficient.  This may in turn lead to significant errors when model predictions are used to interpret experimental observations or to extract interaction parameters. 

The following section presents the model used in this study, which is largely based on the one proposed by~\cite{Korn2007,Korn2008}. This is followed by a description of the simulation algorithm. Section~\ref{s:results} presents a comparison of predictions for high values of $\Pe$ obtained in simulations with and without thermal fluctuations, demonstrating that fluctuations are important for low receptor densities even at very large $\Pe$. A simple model is proposed to explain changes in the state diagram. The effect of HI is then explored by systematically turning off different contributions to HI, with our results suggesting that the choice of simpler descriptions of HI in adhesive dynamics simulations depends on the value of $\Pe$.

\section{\label{s:model} The model}

\begin{figure}[t]
\begin{center}
\resizebox{\columnwidth}{!}{\includegraphics{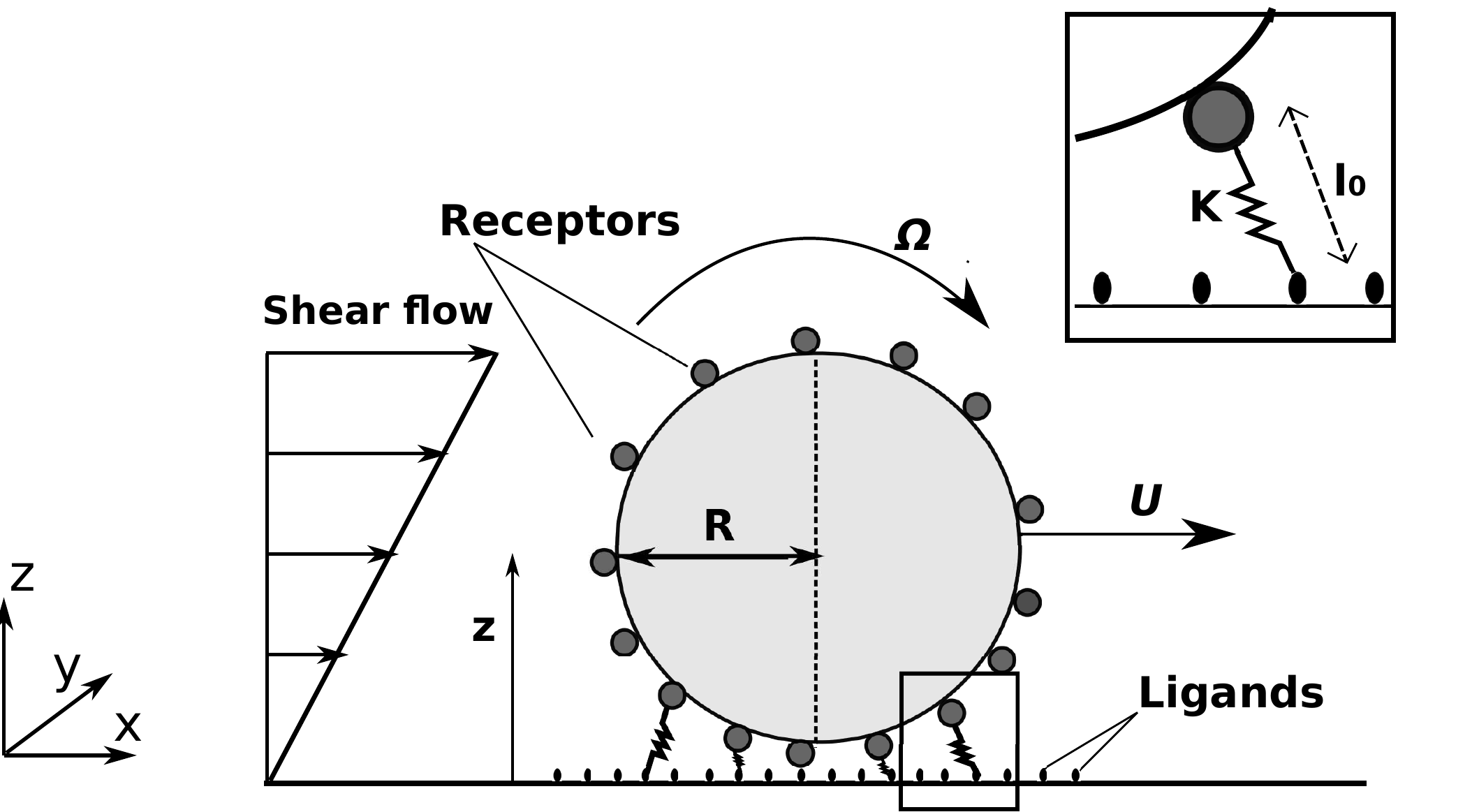}}
\caption{Model schematic}
\label{fig:scheme}
\end{center}
\end{figure}

We simulate the dynamics of spherical particles in a horizontal shear cell with the bottom wall coated by ligands that specifically bind to receptors on the particle surface (Fig. \ref{fig:scheme}). A microparticle is modeled as a rigid sphere of radius $R$. Receptors are modeled as $N_r$, localised reactive patches randomly distributed on the sphere surface each with a spherical capture range of radius $r_0$. Ligands are stationary points on the planar wall distributed on a square grid with spacing $d$ in each direction.Each receptor can form only a single bond with any ligand. Bonds are modelled as semi-harmonic springs with spring constant $\kappa$ and rest length $l_0$. 

Let $\mathbf{X}=[x,\quad y,\quad z,\quad \theta_x, \quad \theta_y, \quad \theta_z]^T $ denote the six-dimensional state vector in which the first three components are the Cartesian coordinates of the center of the sphere  with respect to a fixed frame of reference, and the latter three  describe rotation that maps sphere centred coordinate system and the orientation of the sphere to the laboratory fixed coordinate system. The motion of the particle suspended in a fluid in shear flow near a wall is governed by the following set of coupled It\^{o} stochastic differential equations (SDEs) that account for the translational and rotational Brownian motion of  the particle (\cite{Korn2007, Korn2008}):
\begin{multline}
d\,\mathbf{X}=\left[ \mathbf{U}^{\infty}+ \mathbf{M} \cdot (\mathbf{F}^D + \mathbf{F}^S) +\kBT \, \bm{\nabla}_\mathbf{x} \cdot \mathbf{M} \right]\,d t  \\
+ \sqrt{2\,\kBT}\,\mathbf{B} \,\cdot \,d \,\mathbf{W}_{\!t} \,.
\label{e:langevin}
\end{multline}
Here, $k_B$ is the Boltzmann constant, and $T$ is the absolute temperature, and $\mathbf{U}^{\infty}=[\dot{\gamma}z \quad 0 \quad 0 \quad 0 \quad \dot{\gamma}/2 \quad 0]^T$ is a 6-dimensional  vector containing the translational and rotational unperturbed fluid velocity at the centre of the particle in the absence of the sphere. Wall HI affects the system through the grand mobility tensor $\mathbf{M}$, and the shear-induced force $\mathbf{F}^S$: 
\begin{align}
\mathbf{F}^S =
\begin{bmatrix} \bm{\upzeta}^{td} : \mathbf{E}^{\infty} \\ \bm{\upzeta}^{rd} : \mathbf{E}^{\infty}  \end{bmatrix} \,,
\end{align} 
where $\bm{\upzeta}^{td}$ and $\bm{\upzeta}^{rd}$ are third-rank tensors accounting for the dipolar contributions to translational and rotational  friction, respectively, and
\begin{align}
\mathbf{E}^{\infty}=\begin{bmatrix} 
0 &  0 & \dot{\gamma} \\
0  & 0 & 0 \\
\dot{\gamma} & 0 & 0
\end{bmatrix} \,,
\end{align}
is the rate-of-strain tensor for a homogeneous shear flow of strain-rate $\dot{\gamma}$.  Thermal fluctuations are accounted for by $\mathbf{W}_{\!t}$, which represents a 6-dimensional Wiener process \citep{Gardiner2003, ottbk}. The tensor $\mathbf{B}$ is chosen such that the fluctuation-dissipation theorem is satisfied and $\mathbf{B}\cdot \mathbf{B}^T=\mathbf{M}$. The positional dependence of the mobility tensor leads to the additional drift term proportional to $\bm{\nabla}_{\mathbf{x}} \cdot \mathbf{M}$ which is necessary to ensure that the governing It\^{o} SDEs (Eq.~\eqref{e:langevin}) are consistent with a Fokker-Planck equation for the probability density of the particle position and orientation \citep{Gardiner2003, ottbk}. The tensors $\mathbf{M}$,  $\bm{\upzeta}^{td}$ and $\bm{\upzeta}^{rd}$ and $\mathbf{B}$ depend non-linearly on the distance of the particle centre from from the wall $z$ and its radius $R$. Analytical approximations of these tensors for implementation in Stokesian dynamics 
simulations of a single sphere near a wall have been derived in Ref.~\cite{Cichocki1998273}. These have been summarized previously by \cite{Korn2007} and are also presented here in Appendix~\ref{s:mobility}. 

The vector $\mathbf{F}^D$ denotes the sum of all direct conservative forces and resulting torques acting on the sphere. The first contribution to this generalized force vector comes from a constant attractive  force $\mathbf{F}^G=[0\, 0 \,F^G \,0 \,0\, 0]^T$  towards the wall. The primary role of this force in our simulations is to ensure a steady-state distribution for the particle to be at height $z$ to have a peak at the wall. The second contribution to $\mathbf{F}^D$ comes from the weak non-covalent bonds formed between  receptors and ligands. Each receptor-ligand bond is modeled as a semi-harmonic spring, and  its contribution  to the total force $\mathbf{F}^D$ depends on the instantaneous bond length $r_b$ between centres of the receptor and ligand involved in the bond:
\begin{align}
\mathbf{F}_b = \begin{cases} \kappa 
\,(r_b-\ell_0) \, \hat{\mathbf{r}}_b \,, \quad \textrm{if } r_b > \ell_0 \,, \\
\mathbf{0}\,, \textrm{ otherwise.} 
\end{cases}
\label{e:bforce}
\end{align}
where  $\kappa$ is the spring stiffness,  $\ell_0$ is the rest length of a bond, and $\hat{\mathbf{r}}_b$ is the unit vector from the receptor centre on the sphere surface to the ligand on the wall. In the model proposed by Korn and Schwarz \cite{Korn2008}, $\ell_0$ is not a constant parameter. It is assumed that the bond force $\mathbf{F}_b$ is zero at the instant it is formed. Therefore, $\ell_0$ is set equal to the value of $r_b$ when each bond is formed. The torque exerted by this force on the sphere is  evaluated as  $\mathbf{T}_b = \mathbf{\bar{r}} \times \mathbf{F}_b$, where $\mathbf{\bar{r}}$ is the position vector of that receptor on the sphere surface relative to the sphere centre. 

Ligands are placed on the wall as a periodic square lattice with spacing $d$; in other words, the ligand density on the wall scales as $1/ d^2$. Receptors are distributed randomly on the sphere surface while ensuring that no two receptors overlap within a radius of $r_0$. The receptor density is $N_r/ (4 \pi R^2)$, where $N_r$ is the total number of receptors on the sphere surface. We assume that the capture radius for bond formation is of the same order as receptor size; that is, a receptor-ligand bond can form when a pair is within a distance of $r_0$ from each other. Bond formation is assumed to follow standard reaction kinetics with a fixed average rate-constant $k_\mathrm{on}$.

The kinetics of receptor-ligand dissociation can be more complex. In the simplest case, bonds may break up with a fixed rate constant $k_\mathrm{off}$. More realistically however dissociation is known to depend on the stretching force exerted on a bond. This may in some cases accelerate the break-up (``slip" bonds ) or may strongly slow it down (``catch" bonds), or may result in a combination of both behaviours depending on the strength of the force \cite{Thomas2008}.  Here, following Korn and Schwarz \citep{Korn2008}, we assume slip bonds where each bond dissociates exponentially faster with bond tension. The mean rate is given by Bell's equation \citep{Bell1978}:
\begin{gather}\label{eq:koff}
k_\mathrm{off} \,=\,k_\mathrm{off}^0\, \exp( \frac{F_b}{ F_c}) \,,
\end{gather}
where $k_\mathrm{off}^0$ is the unstressed bond dissociation rate, and $F_c$ is a reactive compliance force scale. 

The dimensional model parameters include the viscosity of the ambient fluid $\eta$, its absolute  temperature $T$, the particle radius $R$ and the shear-rate $\dot{\gamma}$. To express equations and parameters in dimensionless form, we choose $R$, $1/\dot{\gamma}$ and  $6\pi \eta R^2 \dot{\gamma}$ as the characteristic length, time and force scales in the problem. The key dimensionless parameters whose effect we study in our simulations are the P\'{e}clet number,
$\Pe={6\pi\eta R^3 \dot{\gamma}}/{(\kBT)}$, the on-rate $\pi=k_\mathrm{on}/\dot{\gamma}$, the off-rate $\varepsilon_0=k_\mathrm{off}^0/\dot{\gamma}$, and the  receptor density, $N_r$.  Unless otherwise specified, all parameters and variables henceforth will be dimensionless, having been rescaled by the scales given above.  We keep all other parameters fixed at values typical of shear-cell experiments with leukocytes suspended in water at room temperature \citep{Korn2008}: $\Pe=425 \, \, \text{\textemdash} \,\,  42566$; $\pi= 10^{-3} \, \, \text{\textemdash} \, \, 50$; $\varepsilon_0 = 10^{-4}  \, \,\text{\textemdash}\, \, 10^3$; $N_r = 10  \, \, \text{\textemdash}\, \, 5\times 10^3$;  gravitational force, $ F^G = 5 \times 10^{-3}$; receptor size and capture radius, $r_0 = 10^{-2}$; ligand spacing, $d = 5 \times 10^{-2}$; reactive compliance $ F_c =5.3$; bond stiffness, $\kappa =118$. These are obtained by using $R=4.5\mu$m and $\dot{\gamma}=100 \, \, \text{s}^{-1}$ and $\eta=10^{-3}$ Pa s,  relevant to 
leukocytes in an aqueous medium. Ligand density is $1/d^2 = 400$ and the receptor density with $N_r = 10$, and $N_r = 5000$, is 0.796 and 398, respectively. 

The primary observables that help determine the state of motion of a particle near the wall are the average translational velocity in the flow direction $\langle U_x \rangle$ and the average rotational velocity in the flow plane $\langle \Omega_y \rangle$, and their respective variances  $\sigma_U$ and $\sigma_{\Omega}$. Based on their values it is possible to identify distinct states of motion \citep{Korn2008}. If no bonds are formed, the  particle achieves a steady-state average velocity, which we refer to as the hydrodynamic velocity $\langle U_x \rangle_{hd}$. This is the maximum average velocity a particle can attain. When bond formation is insignificant, $\langle U_x \rangle$ with the adhesion kinetics is nearly equal to $\langle U_x \rangle_{hd}$, and the particle is stated to be in a state of ``free-motion".  If bond-formation and disassociation are both significant, the particle can roll at the wall with $ \langle \Omega_y \rangle/\langle U_x \rangle \approx 1$.  When $\Pe \gg 1$, and if bond 
formation dominates over disassociation, the particle is nearly always firmly adhered with the surface, and $\langle U_x \rangle \approx 0$. If bond breakage is a little higher, it is possible to obtain stick-slip motion which is referred to as ``transient adhesion" by \citet{Korn2008}. Table~\ref{tab:table1} summarizes the criteria they suggest for distinguishing between these states. It must be noted that the instantaneous velocities in a stochastic trajectory are not well-defined quantities. The way the averages are calculated is explained in the next section (\ref{s:bds}).

\begin{table}[h]
\caption{\label{tab:table1} Criteria for states of particle motion \citep{Korn2008}}
\centering
\begin{tabular}{cc}
\hline
{\textbf{State}} & {\textbf{Criteria}}\\
\hline
Free motion & $\langle U_x \rangle > 0.95 \,\langle U_x \rangle_{hd}$\\[2mm]

\multirow{2}{*} {Rolling adhesion}  & $\langle \Omega_y \rangle/\langle U_x \rangle > 0.8$ \\
&  and $ 0.95 > \langle U_x \rangle/\langle U_x \rangle_{hd} > 0.01$\\[2mm]

Firm adhesion (FA) & $\langle U_x \rangle < 0.01 \,\langle U_x \rangle_{hd}$ \\[2mm]

\multirow{3}{*} {Transient adhesion I  (TA I)} & $\langle U_x \rangle/\langle U_x \rangle_{hd} > 0.01$ \\
& and $\langle \Omega_y \rangle/\langle U_x \rangle < 0.8$\\
&  and $\sigma_U/\langle U_x \rangle < 0.5$ \\[2mm]

\multirow{3}{*} {Transient adhesion II  (TA II)} & $\langle U_x \rangle/\langle U_x \rangle_{hd} > 0.01$ \\
& and $\langle \Omega_y \rangle/\langle U_x \rangle < 0.8$\\
&  and $\sigma_U/\langle U_x \rangle > 0.5$ \\
\hline
\end{tabular}
\end{table}

\section{\label{s:bds} Simulation algorithm}

The coupled set of SDEs in Eq.~\eqref{e:langevin} are integrated numerically using an Euler discretization \citep{ottbk}. The discrete equation is:
\begin{multline}
\label{eq:euler1}
\Delta \mathbf{X}_t=\left[\mathbf{U}^{\infty}+ \mathbf{M}\cdot (\mathbf{F}^D + \mathbf{F}^S) + \frac{1}{\Pe}\,\bm{\nabla}_\mathbf{x} \cdot \mathbf{M} \right]\,\Delta t  \\ + \sqrt{\frac{1}{\Pe}}\,\mathbf{B} \,\cdot \,\Delta \,\mathbf{W}_{\!t} \,.
\end{multline}	
The Wiener increment $\Delta \mathbf{W}_{\!t}$ in the equation above is a vector with 6 components, each of which is a Gaussian random number with zero mean and variance $2 \Delta t$. The well-known Box-Muller algorithm is first used to transform uniformly distributed random numbers in the interval $(0,1)$ to Gaussian random numbers, which are then multiplied by $ \sqrt{2 \Delta t}$ to generate the Wiener increment.
 
The time-step size $\Delta t$ is chosen to be smaller than all the relevant physical time scales of the system, which are the following: the dimensionless diffusive time scale over which the sphere diffuses in bulk fluid over a distance equal to its own radius is $\Pe$. However, when the sphere is close to the wall and adhesion kinetics are important, the time scale over which the sphere diffuses through a length scale corresponding to the size of a receptor is estimated as $\tau_r=\Pe\, ({r_0/R})^2$. The dimensionless time-scale corresponding to the bond  stiffness is $\tau_f=\kappa^{-1}$. The time scales associated with the on and off-rates are $1/\pi$ and $1/\varepsilon_0$ respectively.

A single simulation for a given set of parameters consists of an ensemble of stochastic particle trajectories. In each trajectory, the initial position of the sphere centre is set as $(x, y, z)=(0, 0, 1 + r_0)$.  Receptor locations are distributed on the sphere surface by randomly (\textit{i.e.} according to a uniform distribution) choosing a set of $N_r$ azimuthal and polar angles in the intervals $[0, \pi]$ and $[0, 2 \, \pi]$, respectively. If the distance between a receptor and any of the previously chosen receptors is less than  $r_0$, the choice is rejected, and a new location is chosen. Once all receptors are located, two separate tables storing the positions of each receptor are created. One stores the coordinates of the position vectors of each receptor in the laboratory-fixed co-ordinate system. The other stores the position vectors  $\mathbf{n}_i$ of each receptor $i$ in a co-ordinate system fixed to the center of the sphere and oriented parallel to the  laboratory-fixed co-ordinate system.

A single time-step in the simulation involved the following sequence of calculations.
\begin{enumerate}
\item Using the position of the centre of the particle, the mobility tensor $\mathbf{M}$ and related quantities $\mathbf{F}^S$, $\bm{\nabla}_\mathbf{x} \cdot \mathbf{M}$ and $\mathbf{B}$ are calculated using the method described by \citet{Korn2007}. This requires the evaluation of several scalar functions of $z$ which are calculated before the start of the simulation and stored in a discrete look-up table. Values at any required $z$ are obtained at each time-step by interpolating between entries in the look-up table.

\item A table of receptors in the contact zone is updated. All receptors for which the arc-length from the lower apex of the sphere is less than $\delta=2 r_0$, are included in the list. Another list of ligands in the contact zone is updated, where the position of these ligands is calculated from the co-ordinates of the sphere centre.

\item The table of bonded receptors and ligands is updated with new bonds. This involves giving every unbonded receptor-ligand pair in the contact zone a chance to form a bond. If the distance between an unbonded  pair is less than $r_0$, the probability that a bond is formed follows Poisson statistics, and is $p_\mathrm{on}=1 - \exp (- \pi\,\Delta t )$ ($\pi$ here is the dimensionless on-rate). Following a Metropolis scheme, a uniform random number is chosen in $[0,1]$; if it is less than $p_\mathrm{on}$, the receptor and ligand are assigned as bonded, and removed from the unbonded list of pairs. The bond distance $r_b$ at this instant is stored in the bond table as the rest length $l_0$ for that bond.

\item The list of bonded pairs is then scanned to calculate all the bond forces (Eq.~\eqref{e:bforce}) and their moments about the particle centre. 
	
\item Each bond is then given a chance to dissociate with a  probability $p_\mathrm{off}=1 -\exp(- k_\mathrm{off}\,\Delta t )$, where the bond-force dependent $k_\mathrm{off}$ is calculated for each bond according to Eq.~\eqref{eq:koff}. Bond dissociation is also implemented following a Metropolis scheme. All dissociated receptors and ligands are removed from the bond table, and their bond forces and torques are set to zero.

\item The total bond force and torque are calculated, and this is added to the gravitational force to obtain the non-hydrodynamic force vector $\mathbf{F}^D$ in Eq.~\eqref{eq:euler1}.

\item The displacement vector $\Delta \mathbf{X}_t$ is calculated according to  Eqn.~\eqref{eq:euler1}, and the position and orientation of the sphere is updated. Wall penetration is prevented by a bounce-back criterion: if  $z (t + \Delta t) < 1$ after the update, it is reassigned as $z (t+\Delta t) = z (t) + | \Delta z | $. An alternative to this strategy is to simply reject the the random vector $\Delta \,\mathbf{W}_{\!t}$ generated if the wall is penetrated and generate a new random vector. This algorithm however requires small time steps to avoid frequent rejections. We find that the bounce-back algorithm gives identical results as the rejection algorithm but is computationally much more efficient. 

\item The receptor location table containing $\mathbf{n}_i$ is updated using the Rodrigues formula:
\begin{gather}
\mathbf{n}_i (t + \Delta t) = \mathbf{n}_i (t) \cos \theta + (\hat{\bm{\theta}} \times \mathbf{n}_i)\, \sin \theta + \hat{\bm{\theta}} \, (\hat{\bm{\theta}} \cdot \mathbf{n}_i) (1-\cos \theta) \,,
\end{gather}
where $\bm{\theta} = (\Delta X_4,\Delta X_5,\Delta X_6)$, $\theta=|\bm{\theta}|$, and  $\hat{\bm{\theta}}=\bm{\theta}/\theta$. Following this, the receptor locations in the laboratory-fixed frame is updated by adding the sphere centre position vector to each $\mathbf{n}_i$.
\end{enumerate}

Each trajectory is allowed to equilibrate for a long time before sampling particle position and orientation co-ordinates and the number of extant bonds  at regular intervals. The length of the equilibration time required to achieve a stationary distribution varies with bond kinetic parameters. As will be shown later, an estimate of this time-scale can be derived. Using these estimates and other standard tests, we ensure that true steady states are obtained.  The values of $\langle U_x\rangle$ and $\langle \Omega_y \rangle$ and their variances are calculated from net displacements between sampling intervals. Averages at each sampling time are calculated as ensemble averages over a large number of independent trajectories. For each $\Pe$  value, the hydrodynamic velocity $\langle U_x\rangle_{hd}$ is determined in a simulation with all bond interactions turned off.

\section{\label{s:results} Results and discussion}
\begin{figure}[t]
\begin{center}
\resizebox{\columnwidth}{!}{\includegraphics{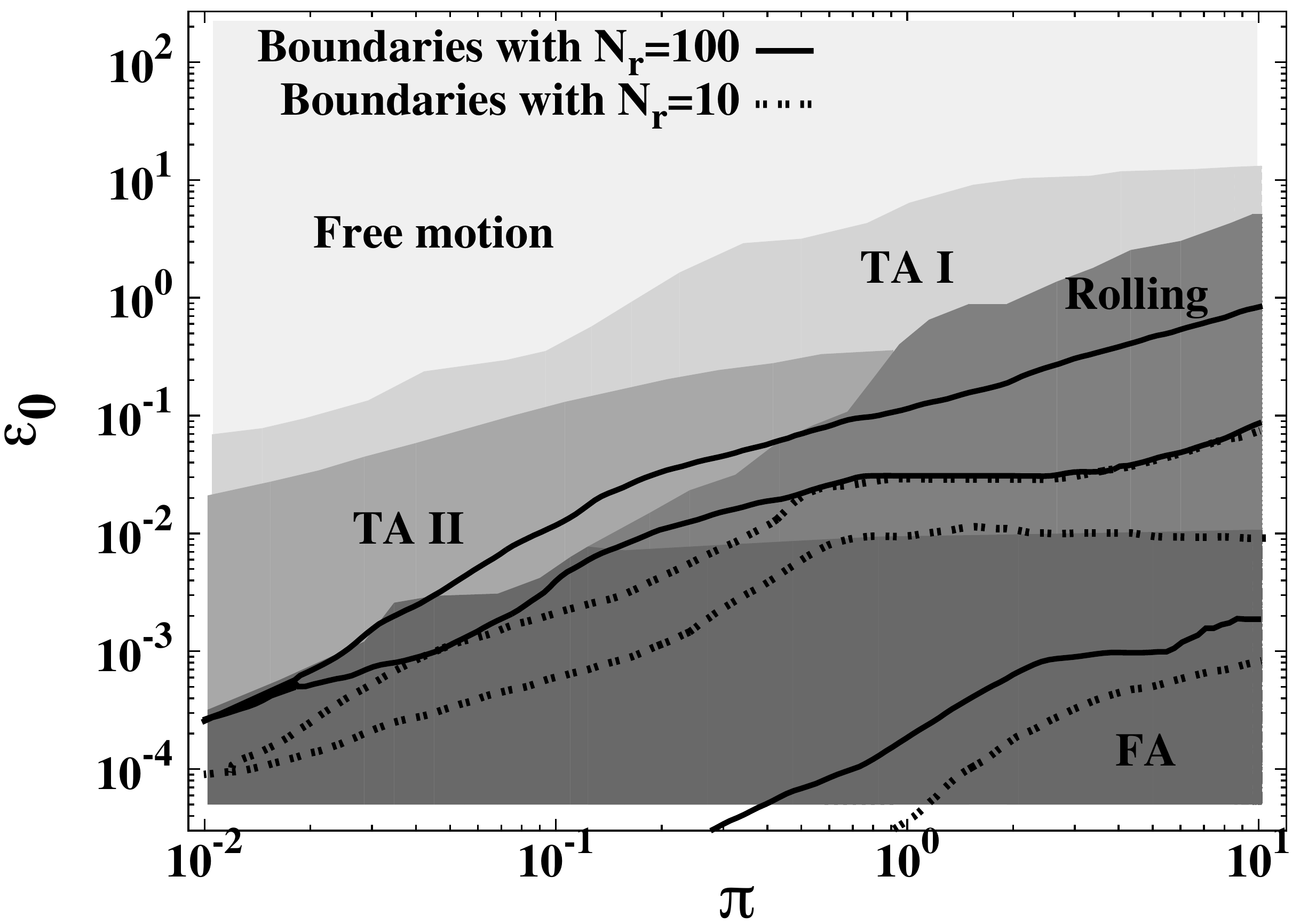}}
\caption{State diagram showing five different dynamic states of motion at $\Pe=42566$: filled areas represent the motion states with $N_r$ = 5000; thick lines are the boundaries between the states of motion with $N_r = 100$ and dotted lines are with $N_r=10$.
}
\label{fig:mys}
\end{center}
\end{figure}

Keeping all the parameters fixed and varying the bond on and off rates $\pi$, and $\varepsilon_0$, the average velocities and their variances
are calculated and a state diagram  is obtained. Figure \ref{fig:mys} maps the space of on- and off-rates showing the five distinct states of motion (as defined by the criteria Table \ref{tab:table1}) that are obtained at a high P\'{e}clet number of $42566$. The boundaries between the various regimes depend on the number of receptors.  When $N_r$ is reduced to 100 and 10, from $N_r$=5000 (with all the other parameters kept the same), two significant trends are observed.   Firstly, there is a shift of all boundaries from the upper left region of the state diagram for the $N_r $= 5000 case towards the lower right region, effectively indicating freer motion of the particle as the chance to form bonds with the wall is increasingly reduced. Secondly, there is a progressive disappearance of the rolling state and its conversion into the transient II regime. Interestingly, we observe that a non-trivial dependence on thermal  fluctuations emerges at low receptor densities. 

\subsection{\label{s:thermal} Significance of thermal fluctuations}
\begin{figure}
\centering
\resizebox{\columnwidth}{!}{\includegraphics{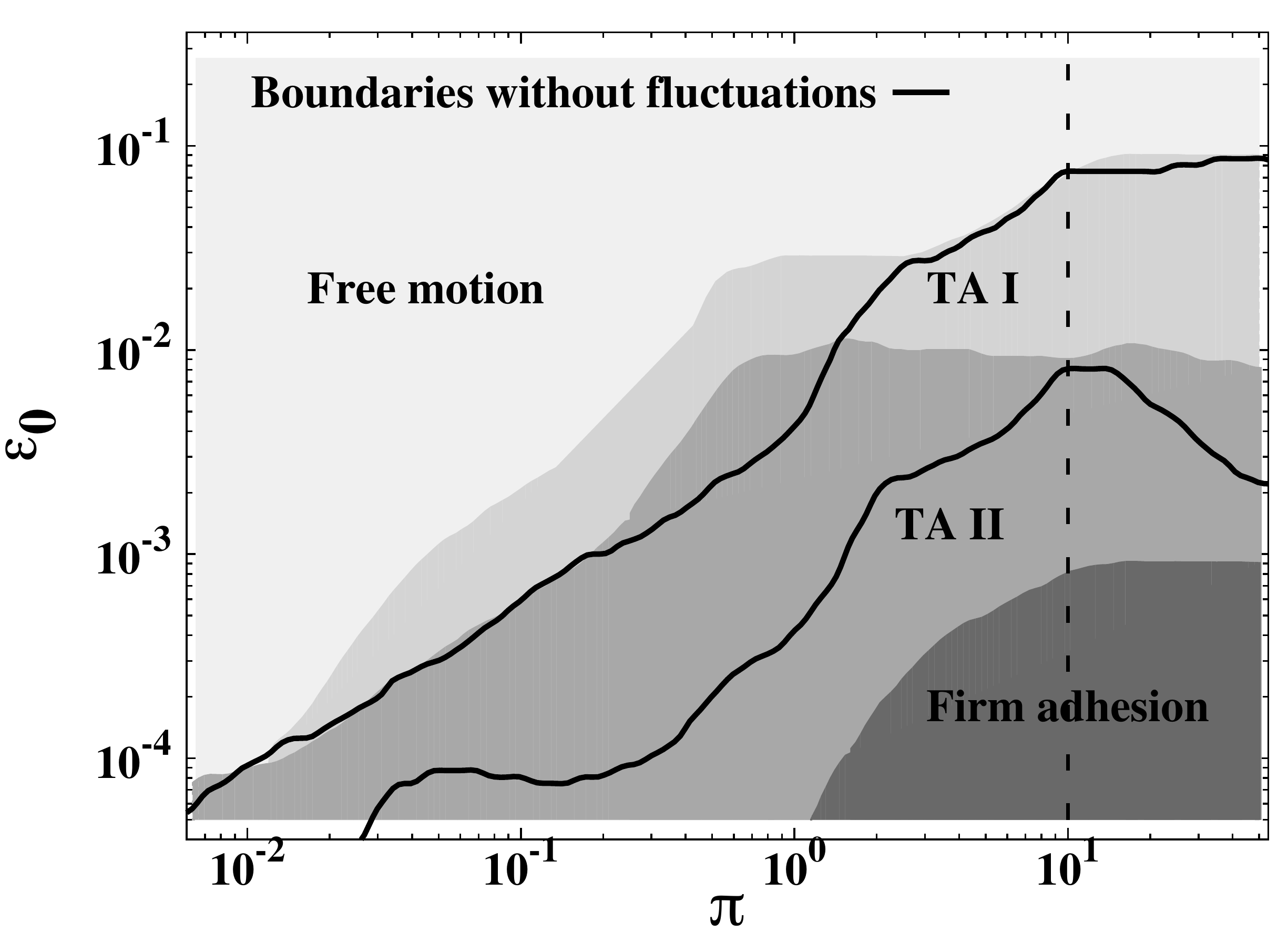}}
\caption{State diagram showing the effect of thermal fluctuations at a low $N_r$ (= 10): filled regions represent states of motion with thermal fluctuations at a high $\Pe$ ( = 42566) and lines are boundaries predicted between regions with the Brownian term turned off. The firm adhesion region is absent without fluctuations. Bond averages and average velocities along the vertical dashed line at $\pi = 10$ are shown in Figs.~\ref{fig:bondsnr10} and~\ref{fig:peVelsnr10}, respectively.} 
\label{fig:mydets10}
\end{figure}

The question we asked is whether a high P\'{e}clet number of 42566 can be regarded as being equivalent to $\Pe = \infty$. In other words, can simulations be run for  large $\Pe$ values be run by simply turning off the Brownian noise term? No significant differences are observed in the state diagrams predicted in simulations with and without fluctuations at high shear rates when receptor numbers are large (at $N_r = 5000$ and 100). The situation changes when $N_r$ is reduced to 10. Motion states with and without fluctuations in this case are displayed in~Fig.~\ref{fig:mydets10}. We observe somewhat counter-intuitively that in the absence of thermal fluctuation, the firm-adhesion region vanishes completely in contrast to simulations with thermal fluctuations. There are also significant shifts in the boundaries the other regions at lower on-rates. 

\begin{figure}[tb]
\begin{center}
\resizebox{\columnwidth}{!}{\includegraphics{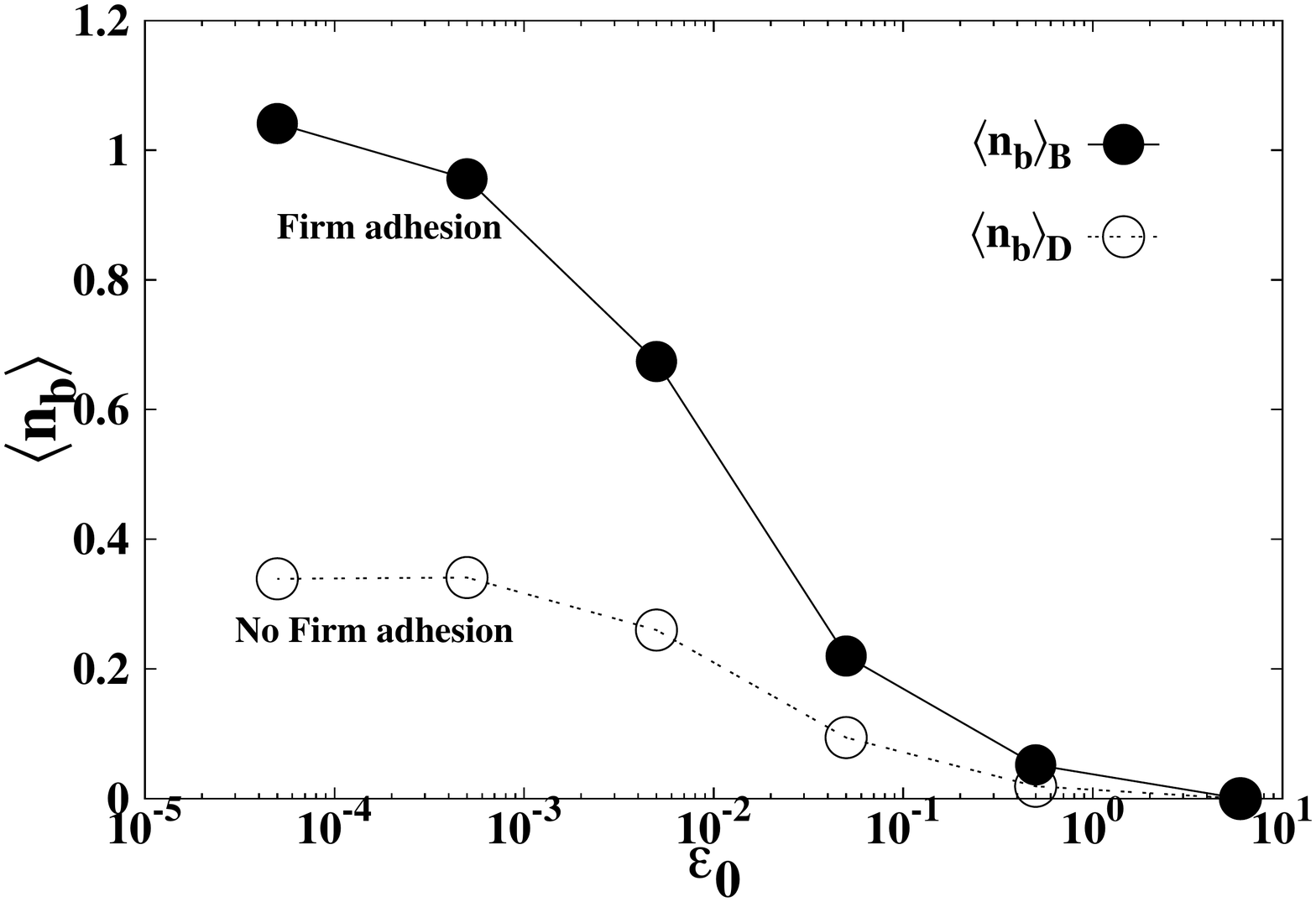}}
\caption{Variation of average number of bonds with the off-rate for $N_r$ = 10 at a fixed on-rate of $\pi$ = 10: subscript B denotes the case with Brownian 
fluctuations, and D denotes without fluctuations.}
\label{fig:bondsnr10}
\end{center}
\end{figure}

For the firm adhesion state to exist, at least one bond must be formed on average so that the particle motion can be completely arrested. In~Fig.~\ref{fig:mydets10},  the average number of bonds is 0.96 -- 1.4 in the firm-adhesion regime when fluctuations are turned on. This is reduced to  0.34 -- 0.48 for the same combination of on- and off-rate when thermal fluctuations are ignored. Figure~\ref{fig:bondsnr10} further shows the variation  of the average number of bonds with the bond off-rate $\varepsilon_0$ at a fixed on-rate of $\pi = 10$ (represented by the vertical dashed line in the state diagram in Fig.~\ref{fig:mydets10}): the average bond number never reaches one in the absence of thermal fluctuations. 
 
\begin{figure}[tb]
\begin{center}
\resizebox{\columnwidth}{!}{\includegraphics{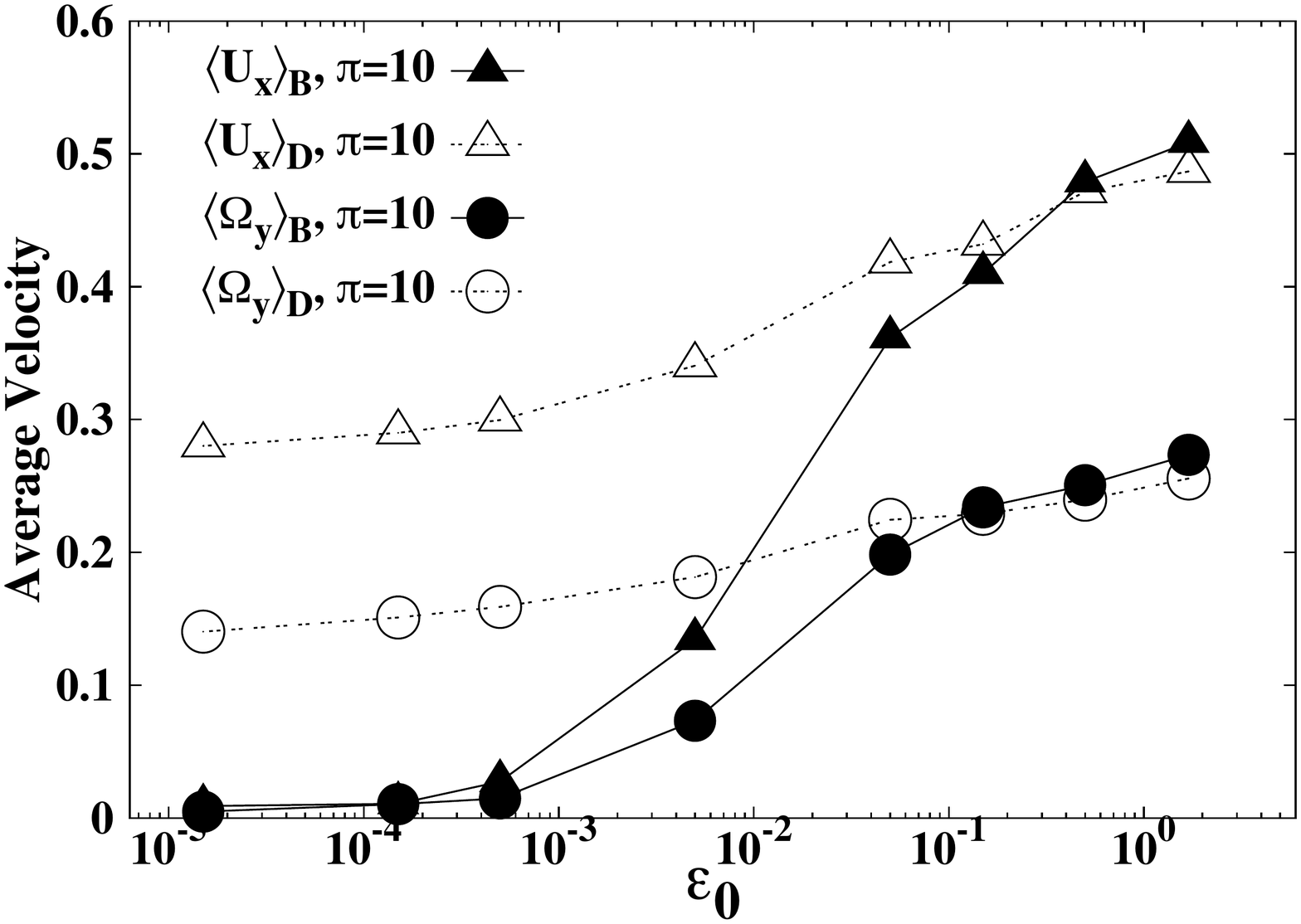}}
\caption{Translational and angular velocities with and without thermal fluctuations at $N_r$ = 10 and $\pi$ = 10: subscript B denotes the case with Brownian 
fluctuations and with D denote without fluctuations.}
\label{fig:peVelsnr10}
\end{center}
\end{figure}

Figure~\ref{fig:peVelsnr10} shows further that the translational as well as angular velocities are considerably reduced to almost zero at low values of $\varepsilon_0$ (at fixed $\pi$) in the presence of thermal fluctuations unlike the case without noise. The expansion of the transient region in Fig.~\ref{fig:mydets10} can be attributed to a fewer number of bonds controlling the dynamics (Fig.~\ref{fig:bondsnr10}). When a single bond is at work, depending upon the off-rate, it can either capture the particle or decelerate it, and the transient II or transient I states are respectively seen. For rolling to occur, whatever the current state of motion is, it should be supported by subsequent formation of bonds. This can happen when there are at least two bonds on average. Figure~\ref{fig:bondsnr10} shows that the average bond number never reaches 2 for both the cases of with and without fluctuations. This explains the absence of rolling in the state diagram at low receptor numbers. 

\begin{figure}[t]
\begin{center}
\includegraphics[width=3in,angle=0]{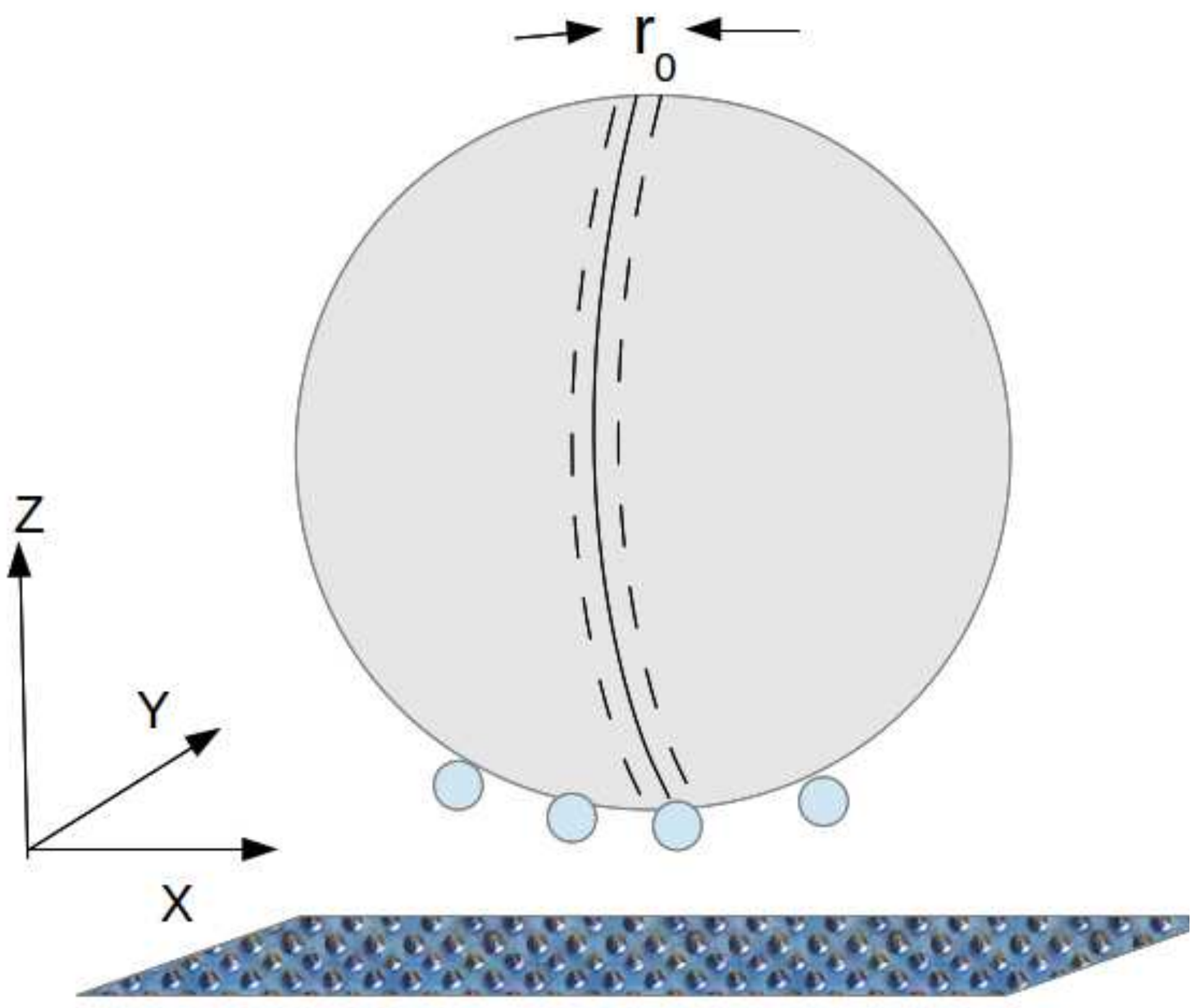}
\caption{Schematic diagram showing a sphere distributed with receptors of radius $r_0$, with the plane of the seam parallel to the $y$-axis.}
\label{fig:cricball}
\end{center}
\end{figure}

When a non-bonded particle is close to the wall, it is forced to spin by the shear flow in the vorticity plane. As pointed out by Korn and Schwarz \cite{Korn2008}, although thermal fluctuations in this plane have little effect on the spinning motion about the vorticity axis at very high shear rates, they also cause rotational diffusion about the other two axes of the particle. The observations above suggest that the effect of this rotational diffusion may facilitate bond formation when the receptor density is very low. A better understanding of the role of thermal fluctuations at low $N_r$ can be obtained with the help of a toy model that is based on a sphere of radius $R$ with just a single receptor and a high density of ligands on the wall, as shown in the schematic diagram in~Fig.~\ref{fig:cricball}. 

In the absence of fluctuations, as the sphere spins along  an axis parallel to the wall with an angular speed of $\omega$, a bond can form only if the receptor \textit{initially} (at the start of the simulation) lies in the seam region of width $r_0$ (the radius of a single receptor) and area $2 \pi r_0 R$. The bond number over all possible initial orientations of favourably or unfavourably placed receptors in a large ensemble of independent particles  is 
\begin{equation}
\langle n_\mathrm{b} \rangle_\textrm{no fluc.}\,  =\, P_\mathrm{seam} \,\overline{n}_\mathrm{b} \,,
\label{e:avgnbdef}
\end{equation}
where $P_\mathrm{seam} (={r_0}/{2R})$ is the probability that the single random receptor is favourably located in the seam initially, and $\overline{n}_\mathrm{b}$ is the time-averaged number of bonds on any single favourable trajectory, which in turn is given by 
\begin{equation*}
 \overline{n}_\mathrm{b} = \frac{\overline{\tau}_{bonded\, |\, seam} }{\overline{\tau}_{unbonded\, |\, seam}  +  \overline{\tau}_{bonded\, |\, seam} } \,,
\end{equation*}
where $ \overline{\tau}$ is the mean time that a favourably located receptor is bonded or unbonded. Estimates for the mean times (derived in Appendix \ref{a:cricmodel}) lead to
\begin{equation}
\langle n_\mathrm{b} \rangle_\textrm{no fluc.}\, = \left(\frac{r_0}{2 R}\right) \, \frac{1 - e^{-k_\mathrm{on} r_0/(R \omega)} }{2 \pi k_\mathrm{off} / \omega\, + \,1 - e^{-k_\mathrm{on} r_0/(R \omega)}}\,.
\label{e:nbnofluc}
\end{equation}

There are two important differences if thermal fluctuations are switched on. Firstly, the ensemble of trajectories is no longer segregated \textit{for all times} into ones with favourable or unfavourable initial conditions. Even if a particle starts out in an unfavourable initial orientation, rotational diffusion about the non-spin axes can bring the receptor into the seam. Equally, receptors on initially favourably oriented particles can also temporarily move out of the seam. This diffusion in and out of the seam continually happens for all trajectories in an ensemble which are now statistically completely equivalent to each other. Secondly, over any single long trajectory, we can distinguish times in which the receptor is either bonded and the particle is still (ignoring the short periods over which a particle is brought to rest after bond formation), or times when it is on the seam but unbonded and the particle is rolling, and times when it is outside the seam and the unbonded particle is rolling.  In 
this case,
\begin{align}
\langle n_\mathrm{b} \rangle\, & =\, \frac{\overline{\tau}_{bonded\, |\, seam} }{\overline{\tau}_{unbonded\, |\, non-seam} + \overline{\tau}_{unbonded\, |\, seam}  +  \overline{\tau}_{bonded\, |\, seam} }
\end{align}
Estimates of the mean times in this case lead to (Appendix \ref{a:cricmodel})
\begin{equation}
\langle n_\mathrm{b} \rangle_\textrm{fluc.} = \frac{1 - e^{-k_\mathrm{on} r_0/(R \omega)}} { {({2 R}/{r_0}) ({2 \pi k_\mathrm{off}}/{ \omega})} + 1 - e^{-k_\mathrm{on} r_0/(R \omega)} } \,.
\label{e:nbfluc}
\end{equation}
Therefore the ratio
\begin{equation}
\frac{\langle n_\mathrm{b} \rangle_\textrm{fluc.}}{\langle n_\mathrm{b} \rangle_\textrm{no fluc.}} = \frac{2 \pi k_\mathrm{off} / \omega\, + \,1 - e^{-k_\mathrm{on} r_0/(R \omega)}} {\displaystyle{ 2 \pi k_\mathrm{off} / \omega\,   + (1 - e^{-k_\mathrm{on} r_0/(R \omega)})\, ({r_0}/{2 R})}} \,.
\label{e:nbratio}
\end{equation}

With or without fluctuations, when  $k_\mathrm{off}$ is large at any fixed $k_\mathrm{on}$, the expressions above lead to the expected result that the mean number of bonds is low, indicating free motion. As $k_\mathrm{off}$ is  decreased however the bond average  $\langle n_\mathrm{b} \rangle_\textrm{fluc.} \rightarrow 1$ whereas $\langle n_\mathrm{b} \rangle_\textrm{no fluc.} \rightarrow r_0/(2 R)$. Therefore, with fluctuations, a transition from free-motion to firm-adhesion should be seen. , whereas in their absence, firm adhesion is never observed since $r_0 \ll R$. Thus, fluctuations about the non-spin axes are essential for binding at low  number of receptors. Figure~\ref{fig:bondsnr10} corroborates this toy model, showing that at lower off-rates, $\langle n_\mathrm{b} \rangle$  approaches 1 in the presence of fluctuation while $\langle n_\mathrm{b} \rangle << 1$ in the absence of fluctuations.

Krobath \textit{et al.} \cite{Krobath2009} discussed another situation where a non-trivial enhancement in bond formation is brought about by thermal fluctuations. Receptors in fluctuating membranes bind to wall ligands in a co-operative manner since changes in flexible membrane conformation when one receptor binds increases the chances that other receptors nearby can also bind. This co-operativity leads to a stronger  sensitivity to receptor and ligand concentrations and potentially larger number of bonds relative to the case where there are no fluctuations and the membrane surface is completely rigid. In the case of a spherical particle however, thermal diffusion enhances the number of bonds by bringing in receptors into the zone where they can bond with ligands; there is no co-operativity involved, since we consider a rigid particle and wall with immobilized receptors and ligands. Further, our analysis suggest that, if the temperature dependence of the on- and off- rate constants is ignored, the relative 
number of bonds formed at steady-state compared to the case of zero fluctuations is independent of temperature itself. Temperature just influences the time scale over which this steady-state is established. At zero temperature, a receptor particle is locked in its initial seam or non-seam location forever. The relative number of bonds formed is temperature dependent n the case of co-operative binding in membranes \cite{Krobath2009}.

\subsection{\label{s:sigHI} Significance of hydrodynamic interactions }
We next turn to understanding which aspects of HI with the wall are most important in determining adhesion dynamics. These wall interactions enter the governing equations in three ways. Firstly, the drag coefficients represented by the diagonal part of the mobility matrix  $\textbf{M}$ sensitively depend upon on the distance  $z$ of the particle from the wall (Fig.~\ref{fig:scheme}). Secondly, HI leads to  a coupling between translational and rotational  degrees of freedom through the non-diagonal terms of $\textbf{M}$. A further third contribution arises from the shear-induced dipole which is represented by the force $\textbf{F}^S$. We examine each of these effects of HI by selectively switching off appropriate terms in the simulations (while keeping thermal fluctuations on). In simulations with no HI, the mobility matrix is position independent, specifically, the diagonal terms are ${1}/{(6 \pi \eta R)}$ and ${1}/{(8 \pi \eta R^3)}$, which correspond to the force and the torque, respectively. The off-
diagonal terms are all zero, and  $\textbf{F}^S$ is also zero.

No significant differences are observed between simulation results of two different flavours of partial HI in  which the $z$-dependent diagonal terms of the mobility matrix are retained:the first, with $\textbf{F}^S$ as well as  the off-diagonal elements of $\textbf{M}$ set to zero, and the second, with the full mobility matrix but with $\textbf{F}^S=0$. There are however clear qualitative differences between the no-HI,  partial-HI with only the diagonal mobility matrix, and the full-HI cases, suggesting that the translation-rotation coupling due to the off-diagonal terms of the mobility matrix is not important under conditions of high shear-rate and large receptor numbers.

\begin{figure}[t]
\begin{center}
\resizebox{\columnwidth}{!}{\includegraphics{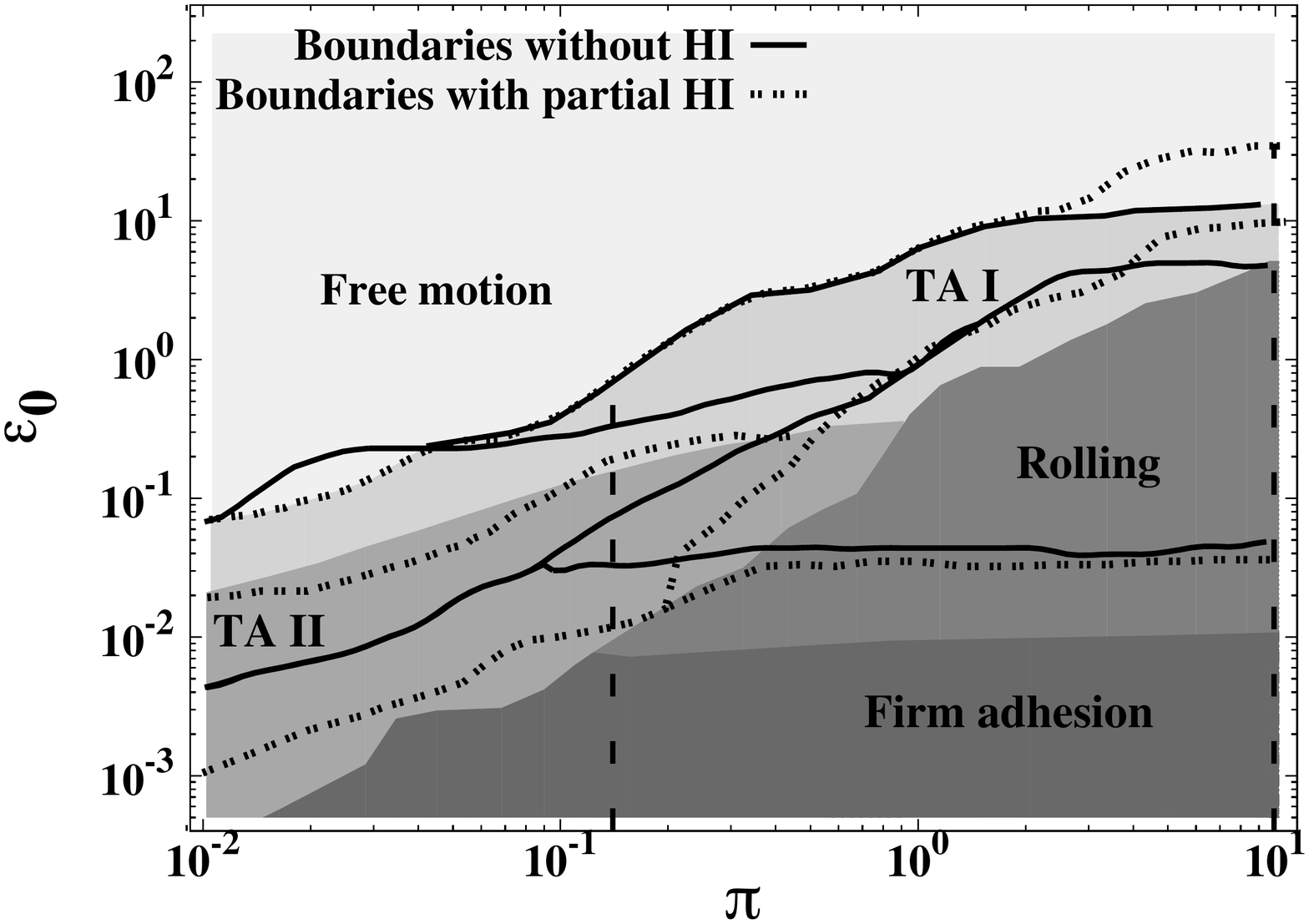}}
\caption{State diagram for the cases of full HI (filled areas), partial HI with only the diagonal terms of the mobility matrix (dotted lines) and no HI 
(thick lines) at $\Pe$ = 42566 and $N_r$=5000: average velocities shown in Fig.~(\ref{fig:Vels}) are at the values of $\pi = 0.15$ and $10$, indicated by the dashed vertical lines at these values of $\pi$.}
\label{fig:HIpHInoHI}
\end{center}
\end{figure}

It can be seen from Fig.~\ref{fig:HIpHInoHI} that HI appears to have little effect on the boundary between free motion the other regimes. It however plays a significant role in the transitions between other states of motion, and the roles of the individual components of HI depend on the on-rate $\pi$. At higher on-rates ($\pi \geq 0.3$), the dominant contribution appears to be that of the shear force; neglecting this means that a larger off-rate is required to observe freer motion at any given on-rate. At lower on-rates --- $0.01 \leq \pi \leq 0.2$ --- as well, neglecting HI has the same qualitative effect, but for these on-rates, the shear-force and the diagonal terms in the mobility matrix appear to contribute equally to the role played by wall HI. Since inertia is negligible, the bond forces are balanced by the total hydrodynamic force, $F^h$. In dimensionless form,  the Bell equation (Eq.~(\ref{eq:koff})) governing the off-rate  is therefore $ \varepsilon = \varepsilon_0 \, \exp {F_h}/{F_c}$. The 
dimensionless hydrodynamic force acting on the sphere with full-HI is approximately $F^h = z/\epsilon + F^S$, where $z$ is the average particle distance from the wall, $\epsilon (< 1)$ is the correction to the drag coefficient due to the presence of wall and arising from the diagonal components of the mobility matrix, and $F^S$ is the shear-induced dipole contribution. In the complete absence of HI, the drag force is smaller and just equal to $F^h = z$. Since we consider particles much larger in size than bond lengths, $z$ does not change much between the no-HI and full HI cases. Therefore, bonds are least stretched when HI is ignored, and most stretched with full HI: that is, HI always favours bond dissociation and requires smaller off-rates to achieve freer states of motion when compared with the no-HI case. On the other hand, larger on rates $\pi$ clearly require larger off rates $\varepsilon$ to a give state of motion different from full adhesion. 
At higher values of $\varepsilon_0$ at which free motion boundary occurs, $\varepsilon_0$ dominates the Bell equation as the shear force and position dependent drag force are not comparatively significant. This results in no perceptible change in the free motion boundary for all the three cases of HI. However at lower $\varepsilon_0$ values, hydrodynamic forces contribution to the force dependent part of Bell equation seems to be significant.

\begin{figure}[t]
\centering
\begin{subfigure}{}
\resizebox{\columnwidth}{!}{\includegraphics{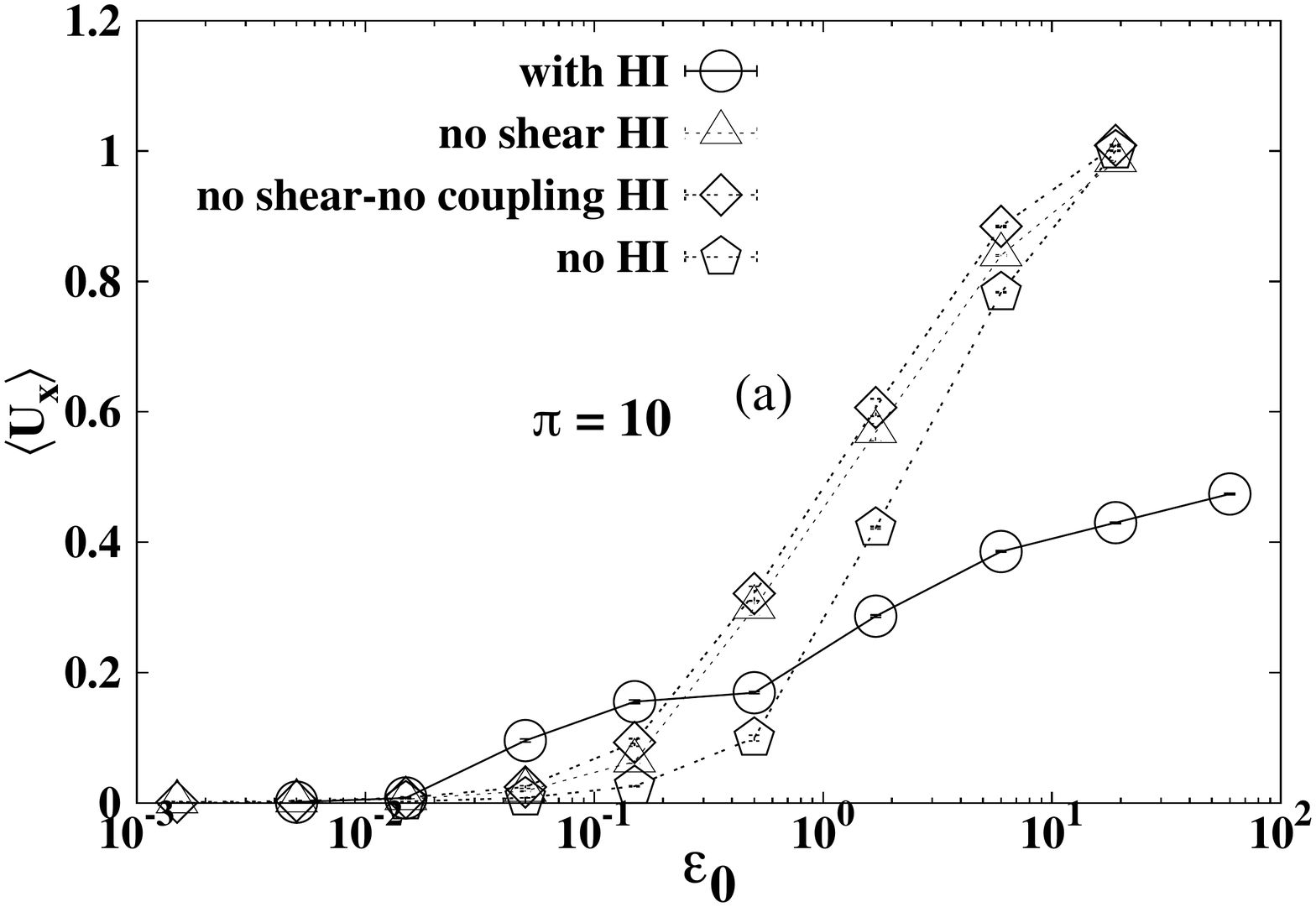}}\\
\end{subfigure}
\begin{subfigure}{}
\resizebox{\columnwidth}{!}{\includegraphics{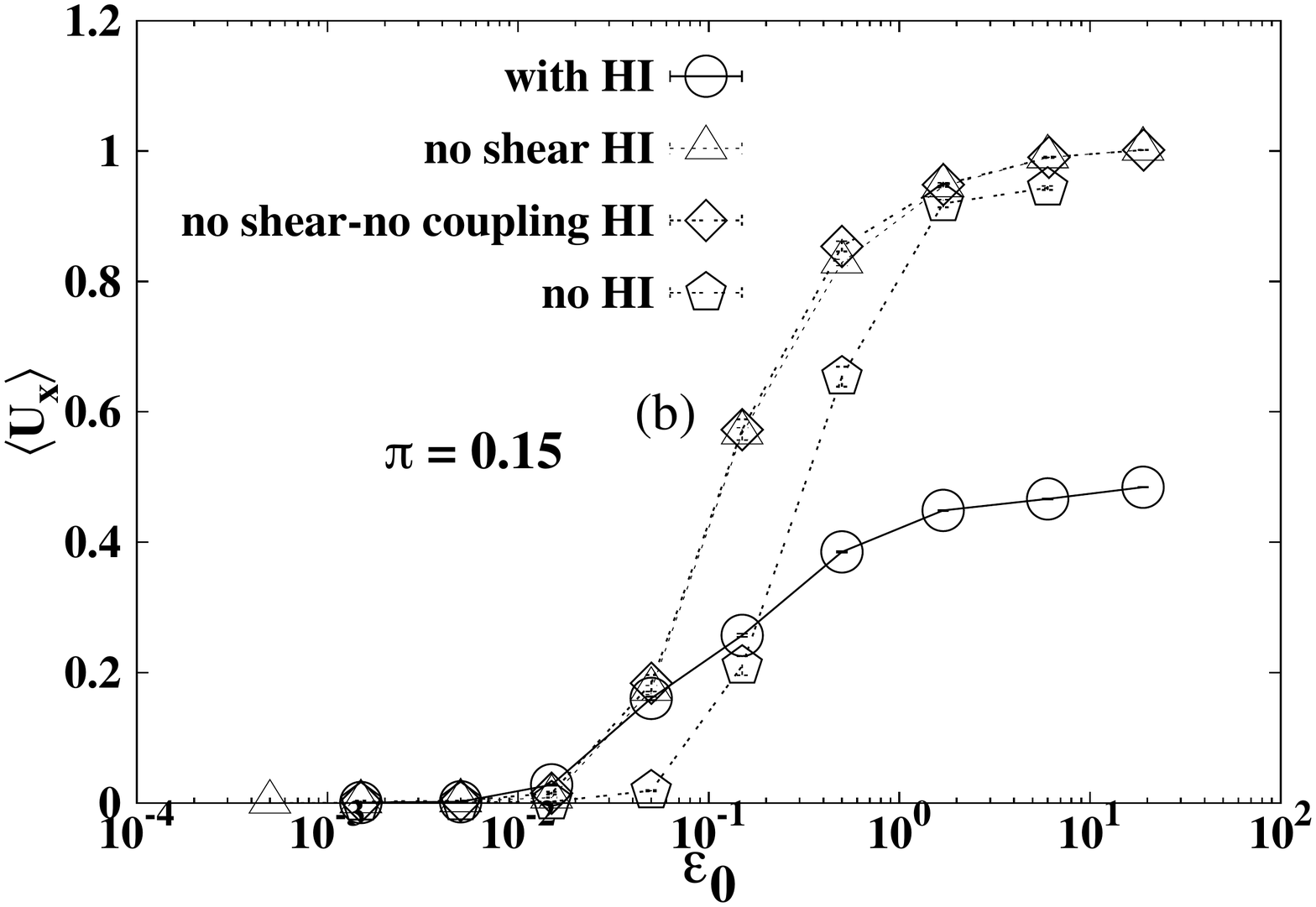}}
\end{subfigure}
\caption{\label{fig:Vels}Average velocities at $\Pe =42566$, $N_r = 5000$ at (a) $\pi = 10$ and (b) $\pi = 0.15$}
\end{figure}

Figure\ref{fig:Vels} compares the effect of HI on the translational velocities at $\pi = 10$ and $\pi=0.15$; similar trends are observed with rotational velocities. From the variation of the velocities with the off-rate $\varepsilon_0$ at fixed $\pi$, it is clear that the shear force is the dominant contribution to the effect of wall HI.  Switching off the off-diagonal terms in the mobility matrix has a relatively minor influence.  Velocities with no-HI and partial-HI are seen to be nearly twice the values in the presence of full-HI. Further it is observed the velocities approach zero at higher off-rates in the case of the no-HI, indicating that bond breakage is assisted by HI. Additionally, at $\pi = 10$, partial-HI and no-HI velocities approach zero at about same value of $\varepsilon_0$. In contrast, at $\pi = 0.15$ they approach zero at  different values of $\varepsilon_0$. This explains why the firm adhesion boundary for the partial-HI and no-HI cases is the same at $\pi = 10$, and different at $\pi = 0.
15$. 
 
\begin{figure}[t]
\begin{center}
\resizebox{\columnwidth}{!}{\includegraphics{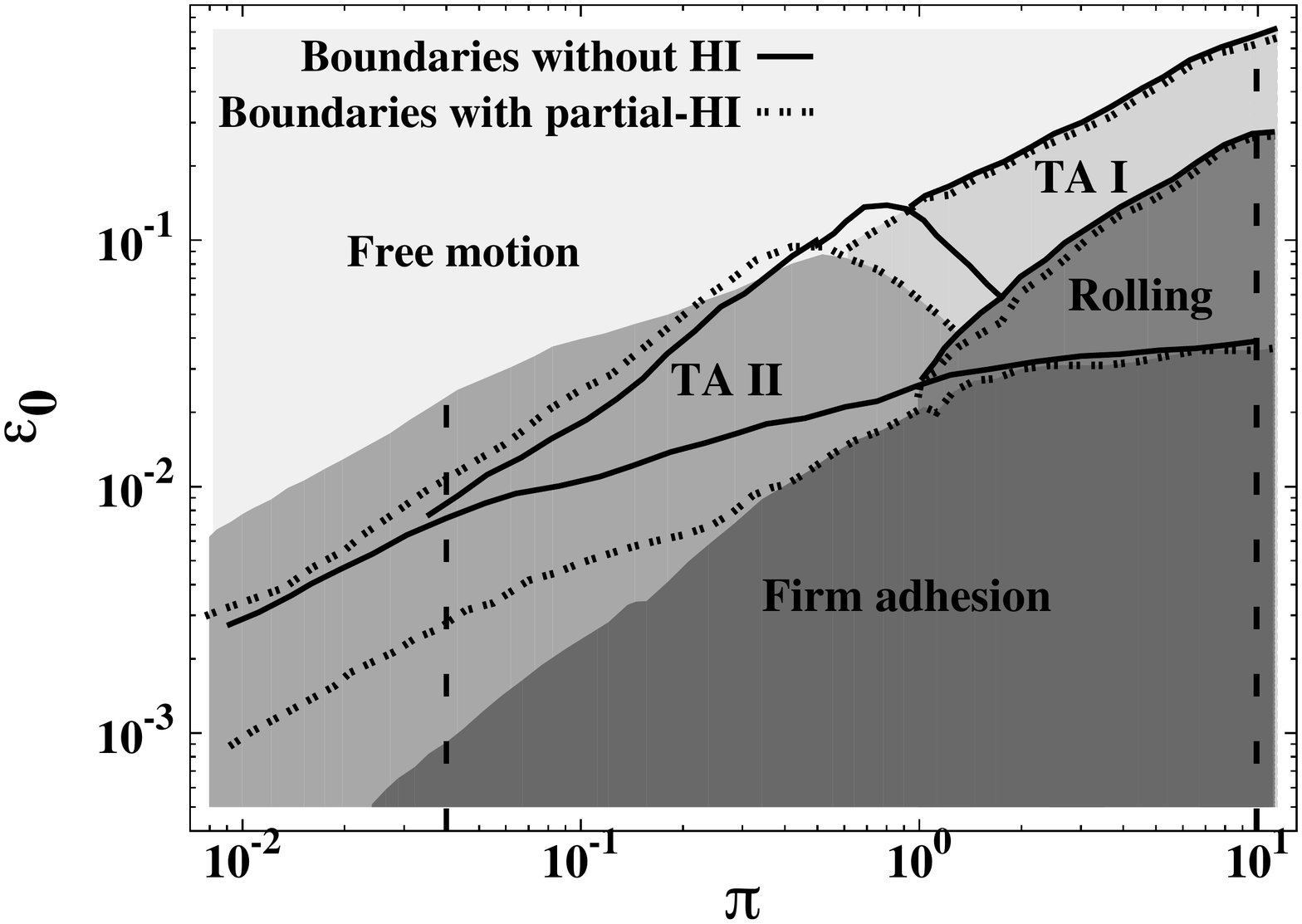}}
\caption{State diagram for the cases of full HI (filled areas), partial HI with only the diagonal terms of the mobility matrix (dotted lines) and no HI 
(thick lines) at $\Pe = 425$ and $N_r = 5000$: average velocities shown in Fig.~(\ref{fig:Vels_2}) are at the values of $\pi = 0.10$ and $0.04$, indicated by the dashed vertical lines at these values of $\pi$.}
\label{fig:HIpe425}
\end{center}
\end{figure} 

\begin{figure}[t]
\centering
\begin{subfigure}{}
\resizebox{\columnwidth}{!}{\includegraphics{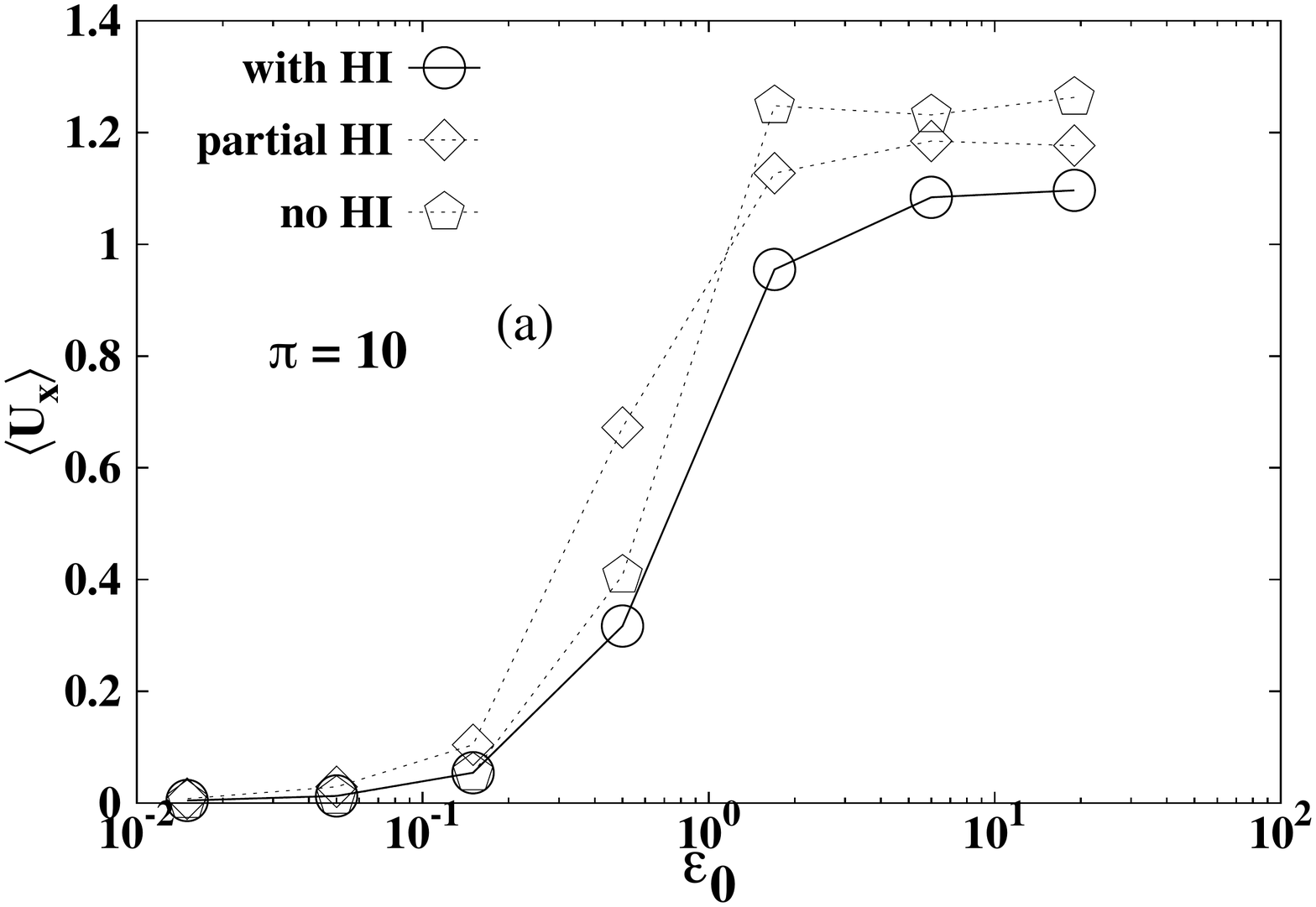}}\\
\end{subfigure}%
\begin{subfigure}{}
\resizebox{\columnwidth}{!}{\includegraphics{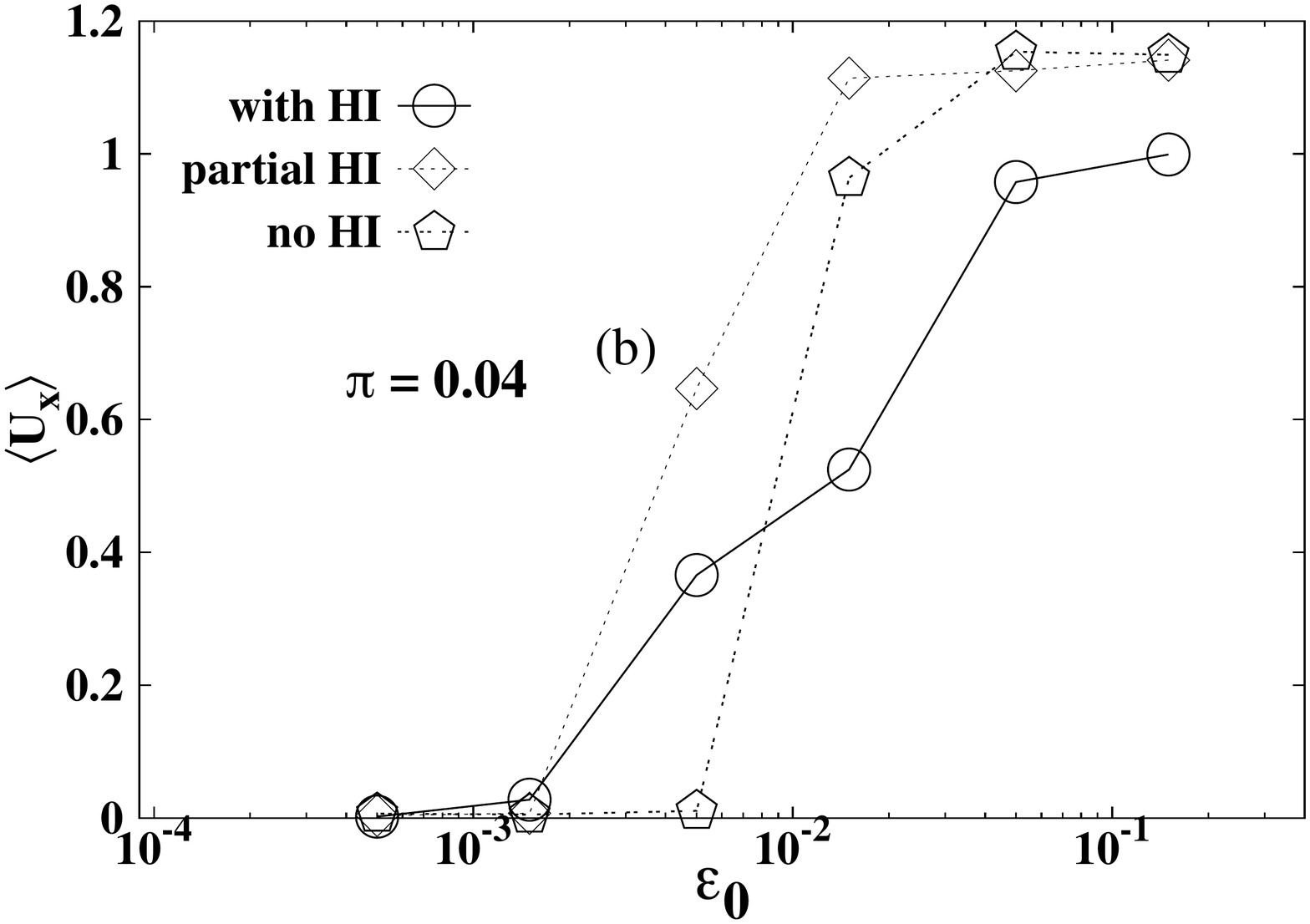}}
\end{subfigure}
\caption{\label{fig:Vels_2}Average velocities at $\Pe =425$ and $N_r = 5000$ at (a) $\pi = 10$ and (b) $\pi = 0.04$ }
 \end{figure}

The results above indicate a strong influence of the shear-induced dipole on the dynamics and motion states of an adhesive particle. In contrast, the state diagram at a lower value of $\Pe=425$ in Fig.~\ref{fig:HIpe425} shows that differences between full-HI, partial -HI and no-HI on the states of motion almost completely vanishes at high on-rates. Figure~\ref{fig:Vels_2} (a) on particle velocities also shows this relatively smaller influence of the shear-dipole force: the plateau values of the velocities at high off-rates (\textit{i.e.} free particle velocities)  at $\pi=10$ are closer for the different HI cases whereas at $\Pe =42566$, turning off the shear-dipole force had a much larger influence. The velocities at $\pi=10$ approach zero at about the same value of $\varepsilon_0$ demonstrating the independence of the state diagram from HI at high values of $\pi$. The effect of the HI however remains quite significant at low values of $\pi$, with the transient-adhesion regions in Fig.~\ref{fig:HIpe425} 
becoming narrower with partial-HI and  disappearing entirely for the no-HI case. The plateau velocity  changes with switching off HI in Fig.~\ref{fig:Vels_2} (b) at low $\pi$ are still relatively small compared to the changes at high $\Pe$. However, the velocity with no HI goes to zero at a higher value of $\varepsilon_0$ compared to the partial-HI and full-HI cases. This is in agreement with the state diagram at low values of $\pi$, which shows that the firm adhesion--transient boundary is progressively shifted upwards, for reasons similar to those discussed earlier at high P\'{e}clet numbers. Notably, the transition from hydrodynamic velocity to zero velocity with decreasing $\varepsilon_0$ is gradual in the case of full-HI, less so with partial-HI, and rapid when there is no HI. This is reflected in the state diagram (Fig. \ref{fig:HIpe425}) as changes in the respective transient adhesion regions. 
 
The BRAD algorithm of English \textit{et al.} \cite{English20043359} completely ignores HI and has been used for simulating the dynamics of viral adhesion. The small size of viruses implies that the P\'{e}clet number for such problems is low. That study was also carried out in the absence of flow. However, we have seen that when the bond on-rates are small, the effect of partial-HI is significant, indicating that the position dependent mobility tensor can affect  bond dynamics. Ignoring the position dependent drag could lead to poor estimates of kinetic parameters when they are extracted by a comparison of simulations with  experimental data. It seems more appropriate to use Eq. (\ref{e:langevin}), with the mobility tensor containing only the diagonal elements (in their full form, with position dependent mobility functions), to obtain a more accurate understanding of adhesive dynamics at low P\'{e}clet numbers. 
 
\section{\label{s:concl} Conclusions}
We have systematically explored the role played by Brownian noise and hydrodynamic interactions in simulations of adhesion of a rigid spherical microparticle to a wall in a shear flow. Particle adhesion in the biological context is of course much more complex. Cells, for instance, are neither spherical nor rigid. For smaller cells and particles such as viruses, receptors and ligands are not simple points of interaction but have a finite size with conformations that are not static. Positions of cell receptors are further not fixed relative to the particle centre but are mobile within the fluid cell membrane. The flexibility of cell membranes is also known to contribute to co-operative binding kinetics. 

Despite the many simplifications made in the model in this study, our conclusions may be expected to be valid quite generally and therefore relevant in more realistic simulations.
Our results firstly show that simulations of microparticles in shear flows must be wary of turning off thermal noise at high shear-rates. Even when the Pecl\'{e}t number is very large, the rotational diffusion out of the vorticity plane caused by thermal fluctuations are particularly important at lower receptor densities and serve to bring receptors into the contact zone for binding with ligands. Switching off thermal fluctuations in simulations at high shear rates with the naive argument that fluctuations are unimportant under such conditions can lead to qualitatively incorrect predictions of various states of motion and the transitions between these with changes in on- and off-rates. Similarly, it is sometimes argued in the interests of simplicity and faster simulations that HI with the wall can be ignored and replaced by just a constant mobility coefficient to obtain qualitative insight.  Our simulations indicate that the role of HI is not just to simply renormalize the friction coefficient. The 
divergence of the friction coefficient as the particle approaches the wall and the shear-induced dipole are observed to be significantly contribute but to differing extents which depend on the on- and off-rates as well as the Pecl\'{e}t number. On the other hand, the translational-rotational coupling in the mobility matrix appears to have little qualitative effect on simulation results. 

Although accurate, the method used here for calculating the mobility matrix are restricted to rigid spheres. Simulations of soft deformable vesicles require coupled solution of the Stokes' flow equations for the fluid and equations governing the dynamics of fluid membranes. Boundary Integral methods \cite{Zhao2011} are highly efficient since they avoid explicit discretization of the Stokes equations for the ambient fluid. Incorporating thermal noise in such simulations is no trivial matter. The Immersed Boundary Method \cite{Pawar2008, Jadhav2007} instead explicitly discretizes the equations for the fluid on an Eulerian grid and the deformable cell on a Lagrangian grid. The method has been extended recently to account for thermal fluctuations \cite{Atzberger2007} but is yet to be applied in the context of deformable particles. Mesoscopic particle methods such as Smoothed Dissipative Particle Dynamics provide a powerful new alternative \cite{Fedosov2014}. Such models can be account for complex receptor 
dynamics on the particle surface  as well as detailed kinetics of interactions with ligands \cite{Thomas2008}.

\appendix
\section{\label{s:mobility} Mobility matrix calculation}
For the sake of completeness, we briefly summarize the theory behind the calculation of the mobility matrix $\mathbf{M}$ for its implementation in Brownian dynamics simulations. A similar  review is also available in Ref.~\cite{Korn2007}. Let $\mathbf{u}^\infty(\mathbf{r})$ be the unperturbed velocity field in the absence of the spherical particle satisfying the no-slip boundary 
condition at the wall. In the shear flow considered here, $u^{\infty}_x = \dot{\gamma} z$ ; $u^{\infty}_y = 0$ ; $u^{\infty}_z = 0$. Let $\mathbf{u} (\mathbf{r})$
be the velocity field that develops in the presence of a spherical particle of radius $R$ and centred at a location $\mathbf{r}_0$ at some instant 
of time and subject to a total non-hydrodynamic force $\mathbf{F}$ and torque $\mathbf{T}$. When particle inertia is negligible, these forces are balanced by
the hydrodynamic force $\mathbf{F}^h$ and torque $\mathbf{T}^h$ on the particle respectively. i.e. $\mathbf{F}=-\mathbf{F}^h$ and $\mathbf{T}=-\mathbf{T}^h$.
In this state of balance, let the translational and rotational velocities of the particle be $\mathbf{U}$ and $\mathbf{\Omega}$ respectively. The velocity field 
$\mathbf{u}(\mathbf{r})$ is such that $\mathbf{u} \rightarrow \mathbf{u}^{\infty}$ as $\mathbf{r} \rightarrow \infty$, and no-slip conditions are satisfied 
at the wall as well as at any position $\mathbf{r}_s$  on the particle surface ($(|\mathbf{r}_s - \mathbf{r}_0| = R$) where the velocity is $\mathbf{u}({\mathbf{r}_s})$ = $\mathbf{U} + \mathbf{\Omega}\times \mathbf{r}_s$. If $\mathbf{f}^h(\mathbf{r}_s)$ is the hydrodynamic traction (force density) due to the fluid stresses on the surface, then from the linearity of the Stokes' equations $\mathbf{f}^h(\mathbf{r}_s)$ and the velocity difference $(\mathbf{u}^{\infty}(\mathbf{r}_s)-\mathbf{u}(\mathbf{r}_s)$ are linearly related to each other. This linear relationship can be expressed most generally as  
\begin{gather}
\mathbf{f}^h(\mathbf{r}_s) = \int_s \, d \mathbf{r}'_s \,\mathbf{Z}(\mathbf{r}_s,\mathbf{r}'_s) \,\cdot\, \left[\mathbf{u}^{\infty}(\mathbf{r}'_s)-\mathbf{u}(\mathbf{r}'_s) \right] \,,
\end{gather}
where $\mathbf{Z}$ is a second-order tensorial function that is independent of either $\mathbf{f}$ or the velocities. For rigid-body rotation of the particle and for a  homogeneous unperturbed velocity field (such as a simple shear flow), 
\begin{align}
\begin{split}
\mathbf{f}^h(\mathbf{r}_s) = \int_s &\, d \mathbf{r}'_s \, \mathbf{Z}(\mathbf{r}_s,\mathbf{r}'_s) \, \cdot\,  \left[ (\mathbf{u}^{\infty}(\mathbf{r}_0) -\mathbf{U} ) \right.  \\ &+ \left. \frac{1}{2} \,\mathbf{E}^{\infty} \cdot (\mathbf{r}'_s - \mathbf{r}_0) + (\mathbf{\Omega}^{\infty} - \mathbf{\Omega}) \times (\mathbf{r}'_s - \mathbf{r}_0) \, \right] \,,
\end{split}
\end{align}
where $\mathbf{E}^{\infty}$ is the rate-of-strain tensor and $\mathbf{\Omega}^{\infty}$ is the angular rotation rate of the unperturbed velocity field. The equation above can in principle be integrated to obtain various surface moments of the traction $\mathbf{f}^h$. It is thus possible to show that the total hydrodynamic force
\begin{align}
\mathbf{F}^h & = \int_s \, d \mathbf{r}_s \,\mathbf{f}^h(\mathbf{r}_s) \,, \notag \\
& = \bm{\upzeta}^{tt} \cdot (\mathbf{u}^{\infty}(\mathbf{r}_0) - \mathbf{U}) + \bm{\upzeta}^{tr} \cdot (\mathbf{\Omega}^{\infty} - \mathbf{\Omega}) + \bm{\upzeta}^{td} \colon \mathbf{E}^{\infty} \,,
\end{align}
and the total hydrodynamic torque,
\begin{align}
\mathbf{T}^h &= \int_s \, d \mathbf{r}_s \,(\mathbf{r}_s - \mathbf{r}_0) \times \mathbf{f}^h(\mathbf{r}_s) \,, \notag \\
&=  \bm{\upzeta}^{rt} \cdot (\mathbf{u}^{\infty}(\mathbf{r}_0) - \mathbf{U}) + \bm{\upzeta}^{rr} \cdot (\mathbf{\Omega}^{\infty}-\mathbf{\Omega}) + \bm{\upzeta}^{rd} \colon \mathbf{E}^{\infty} \,.
\end{align}
Identifying the terms $\bm{\upzeta}^{rd}:\mathbf{E}^\infty$ and $\bm{\upzeta}^{rd}:\mathbf{E}^\infty$ as a shear force and shear torque, we can write 
\begin{align}
 \left[ \begin{array}{cc}
\bm{\upzeta}^{tt} & \bm{\upzeta}^{tr} \\
\bm{\upzeta}^{rt} & \bm{\upzeta}^{rr} \end{array} \right]  \left[ \begin{array}{c}
\mathbf{u}^{\infty}(\mathbf{r}_0) - \mathbf{U} \\
\mathbf{\Omega}^{\infty}(\mathbf{r}_0)-\mathbf{\Omega} \end{array} \right] = \left[ \begin{array}{c}
\mathbf{F}^h - \mathbf{F}^s\\
\mathbf{T}^h - \mathbf{T}^s \end{array} \right] \,,
\end{align}
which then gives the hydrodynamic part of Eq.~\ref{e:langevin}, since
\begin{align}
 \left[ \begin{array}{c}
         \mathbf{U} \\ 
         \mathbf{\Omega}
        \end{array} \right] = 
 \left[ \begin{array}{c}
         \mathbf{u}^{\infty}(\mathbf{r}_0) \\ 
         \mathbf{\Omega}^{\infty}(\mathbf{r}_0)
        \end{array} \right]
+ \mathbf{M} \cdot 
\left[ \begin{array}{c}
\mathbf{F} + \mathbf{F}^s\\
\mathbf{T} + \mathbf{T}^s \end{array} \right] 
\end{align}
where $\mathbf{F} = -\mathbf{F}^h$, $\mathbf{T} = -\mathbf{T}^h$ are the non-hydrodynamic force and torque on the particle as mentioned earlier, and 
\begin{align}
\mathbf{M} =
\left[ \begin{array}{cc} \bm{\upmu}^{tt} & \bm{\upmu}^{tr} \\ \bm{\upmu}^{rt} & \bm{\upmu}^{rr} \end{array} \right] = \left[ 
\begin{array}{cc} \bm{\upzeta}^{tt} & \bm{\upzeta}^{tr} \\ \bm{\upzeta}^{rt} & \bm{\upzeta}^{rr} \end{array}\right]^{-1} \,.
\label{e:mobilities}
\end{align}

The friction tensors $\bm{\upzeta}^{rt}$ \textit{etc.} in the equations above are linear functions of the fluid viscosity but depend in a non-linear fashion on the particle radius and its distance of from the wall.  In the following equations, $\delta_{ij}$ is the Kronecker symbol, and $ \epsilon_{ijk}$ is the Levi-Civita permutation tensor. We denote the flow ($x$), gradient ($z$) and vorticity ($y$) directions as 1, 2 and 3. In simple shear flow \cite{Cichocki1988383,Felderhof198677,Perkins1991575, Schmitz198290}, the friction tensors 
\begin{gather}
\bm{\upzeta}^{tt} = \left(\begin{array}{ccc} \psi^{tt} & 0 & 0\\ 0 & \psi^{tt} & 0\\ 0 & 0 & \phi^{tt} \end{array}\right),\quad \bm{\upzeta}^{rr} = \left(\begin{array}{ccc} \psi^{rr} & 0 & 0\\ 0 & \psi^{rr} & 0\\ 0 & 0 & \phi^{rr} \end{array}\right) \,,\notag \\
\bm{\upzeta}^{tr} = \psi^{tr} \left(\begin{array}{ccc} 0 & 1 & 0\\ -1 & 0 & 0\\ 0 & 0 & 0 \end{array}\right) = (\bm{\upzeta}^{rt})^T\,; \label{e:frictions}
\end{gather}

\begin{gather}
\bm{\upzeta}^{td}_{\alpha} = \left(\begin{array}{ccc} -(1/3)\,\delta_{\alpha 3}\,\phi^{td} & 0 & (1/2)\,\delta_{\alpha 1}\,\psi^{td}\\ 0 & -(1/3)\,\delta_{\alpha 3}\,\phi^{td} & (1/2)\,\delta_{\alpha 2}\,\psi^{td}\\ (1/2)\,\delta_{\alpha 1}\,\psi^{td}& (1/2)\,\delta_{\alpha 2}\,\psi^{td} & (2/3)\,\delta_{\alpha 3}\,\phi^{td} \end{array}\right) \,, \notag \\[2mm] \bm{\upzeta}^{rd}_{\alpha} = \frac{1}{2}\psi^{rd} \left(\begin{array}{ccc} 0 & 0 & \epsilon_{3 \alpha 1}\\ 0 & 0 & \epsilon_{3 \alpha 2}\\ \epsilon_{3 \alpha 1} & \epsilon_{3 \alpha 2} & 0 \end{array}\right)\,; 
\end{gather}

\begin{gather}
\bm{\upzeta}^{dt}_{\alpha} = \left(\begin{array}{ccc} (1/2)\,\delta_{\alpha 3}\psi^{dt} & 0 & (-1/3)\,\delta_{\alpha 1}\,\phi^{dt}\\ 0 & (1/2)\,\delta_{\alpha 3}\,\psi^{dt} & (-1/3)\,\delta_{\alpha 2}\, \phi^{dt}\\ (1/2)\,\delta_{\alpha 1}\,\psi^{dt}& (1/2)\,\delta_{\alpha 2}\psi^{dt} & (-2/3)\,\delta_{\alpha 3}\,\phi^{dt} \end{array}\right)\,, \notag \\[2mm] \bm{\upzeta}^{dr}_{\alpha} = \frac{1}{2}\psi^{dr} \left(\begin{array}{ccc} 0 & \delta_{\alpha 3} & 0\\ -\delta_{\alpha 3} & 0 & 0\\ -\delta_{\alpha 2} & \delta_{\alpha 1} & 0 \end{array}\right)\,. 
\end{gather}

The eight scalar friction functions $\phi^{tt}$, $\psi^{tt}$, $\psi^{tr}$, $\phi^{rr}$, $\psi^{rr}$, $\phi^{td}$, $\psi^{td}$, $\psi^{dr}$ depend on the inverse distance of the sphere from the wall, that is, the dimensionless variable $t = R/ z$, which takes values in the interval [0, 1]. The following eight dimensionless scalar friction functions are first defined:  
\begin{gather*}
\hat{\phi}^{tt} = \phi^{tt}/6\pi \eta R\,, \quad \hat{\psi}^{tt} = \psi^{tt}/6\pi \eta R\,, \quad \hat{\phi}^{rr}   = \phi^{rr}/8\pi \eta R^3 \,,\\
\hat{\psi}^{rr} = \psi^{rr}/8\pi \eta R^3\,, \quad \hat{\psi}^{tr} = \psi^{tr}/8\pi \eta R^2 = -\hat{\psi}^{rt}\,, \\
 \hat{\phi}^{dt} = \phi^{dt}/6\pi \eta R^2 = \hat{\phi}^{td}\,, \quad \hat{\psi}^{dt} = \psi^{dt}/6\pi \eta R^2 = \hat{\psi}^{td}\,, \\ \hat{\psi}^{dr} = \psi^{dr}/8\pi \eta R^3 = -\hat{\psi}^{rd}\,.
\end{gather*}
These are expanded in a Taylors' series in $t$ of the general form,
\begin{equation}
	\hat{\phi} = \sum_{n=0}^{\infty} f_n t^n \,,
	\label{taylor}
\end{equation}
and several coefficients $f_n$ of such series expansions for each of the eight friction functions are tabulated in Refs. \cite{Cichocki1998273,Perkins1991575}. Far away from the wall, the series converge rapidly as $t \rightarrow  0$. Close to the wall, convergence is poor as $t\rightarrow   1$; however, in this limit, analytical results exist from lubrication theory that have the following general form for all scalar friction functions:
\begin{equation}
\hat{\phi}(t) = C_1\, \frac{t}{1-t} + C_2\, \ln(1-t) + C_3 + C_4 \,\frac{1-t}{t}\ln(1-t) + O(1-t) \,.
\label{lub}
\end{equation}
The coefficients $C_1 ,C_2 ,C_3 ,C_4$ for each of the eight friction functions are given in Ref. \cite{Perkins1991575}. For numerical implementation covering the whole range of $t$, solutions at the two limits  are matched using a Pad\'{e} scheme such that,
\begin{equation}
\hat{\phi}(t) = C_1 \frac{t}{1-t} + C_2 \ln(1-t) + C_3 + C_4 \frac{1-t}{t}\ln(1-t) + P_{N}(t)
\label{pade}
\end{equation}
where
\begin{equation}
P_N(t) = \frac{a_0+a_1t+a_2t^2+.....+a_Nt^N}{1+b_1t+b_2t^2+.....+b_Nt^N} \
\end{equation}
is the Pad\'{e} approximant of the first $N$ terms of the Taylors' series expansion $\sum_{n=0}^{N} g_n t^n$ of the difference
\begin{equation}
\sum_{n=0}^{N} f_n t^n - C_1 \frac{t}{1-t} - C_2 \ln(1-t) - C_3 - C_4 \frac{1-t}{t}\ln(1-t) \,.
\label{reslt}
\end{equation}
The coefficients $g_n$ can be calculated from the known coefficients $f_n$ and the $C$-constants. Then, the coefficients of the rational function $P_N$ are obtained as the solution of the following system of linear equations,
\begin{align}
\sum_{n=1}^N b_n g_{N-n+k} = -g_{n+k}\,, \\
\sum_{n=1}^N b_n g_{k-n} - a_k = 0 \,, 
\end{align}
with $k = 1,...,N$.

With all the $C$- and  Pad\'{e} coefficients in hand, the values of the eight dimensionless friction functions can be calculated at any position of the particle from the wall. From these, the value of the following scalar mobility functions are obtained: 
\begin{gather*}
\hat{\alpha}^{tt} = \frac{1}{\hat{\phi}^{tt}}\,, \quad \hat{\beta}^{tt} = \frac{\hat{\psi}^{rr}}{ \hat{\psi}^{tt} \hat{\psi}^{rr} - (4/3) (\hat{\psi}^{tr})^2} \,,\\
\hat{\alpha}^{rr} = \frac{1}{\hat{\phi}^{rr}}\,, \quad \hat{\beta}^{rr} = \frac{\hat{\psi}^{tt}}{ \hat{\psi}^{tt} \hat{\psi}^{rr} - (4/3) (\hat{\psi}^{tr})^2} \,,\\
\hat{\beta}^{tr} = - \frac{4}{3} \, \frac{\hat{\psi}^{tr}}{ \hat{\psi}^{tt} \hat{\psi}^{rr} - (4/3) (\hat{\psi}^{tr})^2} \,,\\
\hat{\alpha}^{dt} = - \hat{\phi}^{dt} \, \hat{\alpha}^{tt}\,,\quad \hat{\beta}^{dt} = - \hat{\psi}^{dt} \, \hat{\beta}^{tt} - \hat{\psi}^{dr} \, \hat{\beta}^{tr} \,,\\
\hat{\beta}^{dr} = -\frac{3}{4}\, \hat{\psi}^{dt} \hat{\beta}^{tr} - \hat{\psi}^{dr} \hat{\beta}^{rr} \,.
\end{gather*}
The mobility tensors $\bm{\upmu}^{tt}$, $\bm{\upmu}^{rr}$, $\bm{\upmu}^{tr}$ and $\bm{\upmu}^{rt}$ that make up the overall mobility matrix $\mathbf{M}$ as shown in Eq.~\ref{e:mobilities} have the same forms as the friction tensors given by Eq.~\ref{e:frictions}, but with $\alpha$  in place of $\phi$, and $\beta$ in place of $\psi$.

\section{\label{a:cricmodel} Derivation for the ratio of average bond number with and without thermal fluctuations}

Consider the case of a single receptor on a sphere of radius $a$. We want to estimate the mean time for first bond formation, when the sphere spins along the $y$-axis parallel to the wall with an angular speed of $\omega$, and firstly,  in the absence of thermal fluctuations. A bond in this case can only form if the receptor lies in the seam region of width $r_0$ and area $2 \pi r_0 R$. The probability that a receptor lies in the seam initially is hence $P_\mathrm{seam}=2 \pi r_0 R/ (4 \pi R^2 )=r_0/(2 R)$. The first-order reaction kinetics for bond-formation essentially implies that bond formation proceeds as a Poisson process with a mean rate $k_\mathrm{on}$. Now if a receptor lies in the seam, then its residence time in the reaction zone of size $r_0 \times r_0$ is $\tau_\mathrm{r}=r_0/ (R \omega)$. Given that the receptor lies on the seam, the conditional probability that adhesion occurs within this residence time, for any single pass of  the receptor across the ligand, is 1 minus the probability that 
no bond forms at all within $\tau_\mathrm{r}$. From the statistics for a Poisson process
\begin{align}
P_\mathrm{bond\, |\, seam} &=1 - \frac{ e^{-k_\mathrm{on} \tau_\mathrm{r}} (k_\mathrm{on} \tau_\mathrm{r})^0}{0!}\,,\notag \\
&=1 - e^{-k_\mathrm{on} r_0/(R \omega)} \,.
\end{align}
If this is the probability that a bond forms in one pass, then the mean number of passes before a bond is formed is  $1/P_\mathrm{bond\, |\, seam}$.  Since we have one pass per revolution, and the time for one revolution is $2 \pi/\omega$, the mean time before first bond formation, or the mean time for which the receptor is unbonded, given a receptor is on the seam is 
\begin{align}
\overline{\tau}_{unbonded\, |\, seam} &=\frac{2 \pi}{\omega \,P_\mathrm{bond\, |\, seam}}\,,\\
&=\frac{2 \pi}{\omega (1 - e^{-k_\mathrm{on} r_0/(R \omega)} )}\,.
\end{align}

Once a bond forms the mean time for it to de-bond, or the mean-time for which the receptor is bonded, is  just $\overline{\tau}_{bonded\, |\, seam}=1/k_\mathrm{off}$, since the spin is switched off once a bond forms. This is based on the distribution of waiting times for a Poissonian event: $\lambda e^{-\lambda t}$. The mean waiting time is $1/\lambda$. In our case, $\lambda = k_\mathrm{off}$.  Since a bond is present only during the time debonding takes place, the time-averaged number of bonds on any single on-seam-receptor trajectory is hence
\begin{align}
\overline{n}_\mathrm{b} &=\frac{\overline{\tau}_{bonded\, |\, seam} }{\overline{\tau}_{unbonded\, |\, seam}  +  \overline{\tau}_{bonded\, |\, seam} }\,\notag \\
&=\frac{1 - e^{-k_\mathrm{on} r_0/(R \omega)} }{2 \pi k_\mathrm{off} / \omega\, + \,1 - e^{-k_\mathrm{on} r_0/(R \omega)}}\,.
\end{align}
Note that in order to reach a steady state, we need to sample over a time scale greater than $\overline{\tau}_{unbonded\, |\, seam}  +  \overline{\tau}_{bonded\, |\, seam}$ which diverges as either $k_\mathrm{on}$ or $k_\mathrm{off} \rightarrow 0$. The result above is the average over the trajectories that start with a favourably-placed receptor. The mean bond number over the entire ensemble, including those initial orientations that can never form a bond is 
\begin{align}
\langle n_\mathrm{b} \rangle_\textrm{no fluc.}\, & =\, P_\mathrm{seam} \,\overline{n}_\mathrm{b}\,,\notag \\
&= \frac{r_0}{2 R} \, \frac{1 - e^{-k_\mathrm{on} r_0/(R \omega)} }{2 \pi k_\mathrm{off} / \omega\, + \,1 - e^{-k_\mathrm{on} r_0/(R \omega)}}\,.
\end{align}

If thermal fluctuations are switched on, there are two important differences. Firstly, the ensemble of trajectories is no longer segregated into ones with favourable or unfavourable initial conditions; all trajectories are statistically equivalent. Secondly, over any single long trajectory, we can distinguish times in which the receptor is bonded, times when it is in the seam and unbonded and times when it is outside the seam and unbonded. In this case, therefore,
\begin{align}
\langle n_\mathrm{b} \rangle_\textrm{fluc.}\, & =\, \frac{\overline{\tau}_{bonded\, |\, seam} }{\overline{\tau}_{unbonded\, |\, non-seam} + \overline{\tau}_{unbonded\, |\, seam}  +  \overline{\tau}_{bonded\, |\, seam} } \,, 
\end{align}
At steady state, the equality of probability fluxes into and out of seam regions implies that the fraction of time that the receptor spends on the seam but unbonded is exactly the same as the fractional area of the seam:
\begin{align}
\frac{\overline{\tau}_{unbonded\, |\, seam}}{\overline{\tau}_{unbonded\, |\, seam}+\overline{\tau}_{unbonded\, |\, non-seam} } =  \frac{r_0}{2 R} \,.
\end{align}
The expressions for  $\overline{\tau}_{unbonded\, |\, seam}$ and $\overline{\tau}_{unbonded\, |\, seam}$ derived earlier are still valid. Therefore, 
\begin{align}
\langle n_\mathrm{b} \rangle_\textrm{fluc.}\, 
= \frac{1 - e^{-k_\mathrm{on} r_0/(R \omega)}} {\displaystyle{\frac{2 R}{r_0}\, \frac{2 \pi k_\mathrm{off}}{ \omega}\,  } + 1 - e^{-k_\mathrm{on} r_0/(R \omega)} } \,.
\label{e:nbfluc1}
\end{align}
Thus,
\begin{align}
\frac{\langle n_\mathrm{b} \rangle_\textrm{fluc.}}{\langle n_\mathrm{b} \rangle_\textrm{no fluc.}} 
& = \frac{2 \pi k_\mathrm{off} / \omega\, + \,1 - e^{-k_\mathrm{on} r_0/(R \omega)}} {\displaystyle{ 2 \pi k_\mathrm{off} / \omega\,   + (1 - e^{-k_\mathrm{on} r_0/(R \omega)})\, \frac{r_0}{2 R}}} \,.
\end{align}


\end{document}